\documentclass[11pt]{article}
\usepackage{geometry}
\usepackage[ruled,vlined]{algorithm2e}
\usepackage{amsmath}
\usepackage{amssymb}
\usepackage{bbold}
\usepackage{appendix}
\usepackage{balance}
\usepackage{caption}
\usepackage{color}
\usepackage{comment}
\usepackage{epstopdf}
\usepackage{graphicx}
\usepackage{listings}
\usepackage{mathtools}
\usepackage{microtype}
\usepackage{multirow}
\usepackage{pdfpages}

\usepackage{subfig}
\usepackage{tabularx}
\usepackage{times}
\usepackage{url}
\usepackage{xcolor}
\usepackage{xspace}
\usepackage{bm}
\usepackage{mdwlist}
\usepackage{hyperref}

\usepackage{breqn}

\newcommand{\bloomfilter}{Bloom filter\xspace}
\newcommand{\bloomfilters}{Bloom filters\xspace}
\newcommand{\hashmap}{Hash-map\xspace}
\newcommand{\hashmaps}{Hash-maps\xspace}
\newcommand{\btree}{B-Tree\xspace}
\newcommand{\btrees}{B-Trees\xspace}
\newcommand{\lookup}{look-up\xspace}
\newcommand{\lookups}{look-ups\xspace}
\newcommand{\myemail}[1]{\tt #1}
\newcommand{\naive}{na\"{\i}ve\xspace}
\newcommand{\Naive}{Na\"{\i}ve\xspace}

\newcommand{\framework}{{\sc LIF}}

\begin{document}
\title{The Case for Learned Index Structures}

\author{
Tim Kraska\footnote{Work done while author was affiliated with Google.}\\
MIT\\
Cambridge, MA\\
\myemail{kraska@mit.edu}
\and 
Alex Beutel\\
Google, Inc.\\
Mountain View, CA\\
\myemail{alexbeutel@google.com}
\and 
Ed H. Chi\\
Google, Inc.\\
Mountain View, CA\\
\myemail{edchi@google.com}
\and 
Jeffrey Dean\\
Google, Inc.\\
Mountain View, CA\\
\myemail{jeff@google.com}
\and 
Neoklis Polyzotis\\
Google, Inc.\\
Mountain View, CA\\
\myemail{npolyzotis@google.com}
}
\date{}

\maketitle

\begin{abstract}
Indexes are models: a \btree-Index can be seen as a model to map a key to the position of a record within a sorted array, a Hash-Index as a model to map a key to a position of a record within an unsorted array, and a BitMap-Index as a model to indicate if a data record exists or not. 
In this exploratory research paper, we start from this premise and posit that all existing index structures can be replaced with other types of models, including deep-learning models, which we term {\em learned indexes}.
The key idea is that a model can learn the sort order or structure of lookup keys and use this signal to effectively predict the position or existence of  records.
We theoretically analyze under which conditions learned indexes outperform traditional index structures and describe the main challenges in designing learned index structures. 
Our initial results show, that by using neural nets we are able to outperform cache-optimized \btrees by up to $70\%$ in speed while saving an order-of-magnitude in memory over several real-world data sets. 
More importantly though, we believe that the idea of replacing core components of a data management system through learned models has far reaching implications for future systems designs and that this work just provides a glimpse of what might be possible. 
\end{abstract}

\section{Introduction}
\label{sec:intro}
Whenever efficient data access is needed, index structures are the answer, and a wide variety of choices exist to address the different needs of various access patterns.
For example, \btrees are the best choice for range requests (e.g., retrieve all records in a certain time frame); 
\hashmaps are hard to beat in performance for single key look-ups; and \bloomfilters are typically used to check for record existence. Because of their importance for database systems and many other applications, indexes have been extensively tuned over the past decades to be more memory, cache and/or CPU efficient \cite{survey_btree,survey_hash,bloom1,bloom2}.

Yet, all of those indexes remain general purpose data structures; they assume nothing about the data distribution and do not take advantage of more common patterns prevalent in real world data.
For example, if the goal is to build a highly-tuned system to store and query ranges of fixed-length records over a set of continuous integer keys (e.g., the keys 1 to 100M), one would not use a conventional \btree index over the keys since the key itself can be used as an offset, making it an $O(1)$ rather than $O(\log n)$ operation to look-up any key or the beginning of a range of keys. 
Similarly, the index memory size would be reduced from $O(n)$ to $O(1)$.
Maybe surprisingly, similar optimizations are possible for other data patterns.
In other words, knowing the exact data distribution enables highly optimizing almost any index structure.  

Of course, in most real-world use cases the data do not perfectly follow a known pattern and the engineering effort to build specialized solutions for every use case is usually too high. 
However, we argue that machine learning (ML) opens up the opportunity to learn a model that reflects the patterns in the data and thus to enable the automatic synthesis of specialized index structures, termed {\bf learned indexes}, with low engineering cost.

In this paper,  we explore the extent to which learned models, including neural networks, can be used to enhance, or even replace, traditional index structures from \btrees to \bloomfilters. 
This may seem counterintuitive because ML cannot provide the semantic guarantees we traditionally associate with these indexes, and because the most powerful ML models, neural networks, are traditionally thought of as being very compute expensive. 
Yet, we argue that none of these apparent obstacles are as problematic as they might seem. 
Instead, our proposal to use learned models has the potential for significant benefits, especially on the next generation of hardware.

In terms of semantic guarantees,  indexes are already to a large extent learned models making it surprisingly straightforward to replace them with other types of ML models. 
For example, a \btree  can be considered as a model which takes a key as an input and predicts the position of a data record in a sorted set (the data has to be sorted to enable efficient range requests). 
A \bloomfilter is a binary classifier, which based on a key predicts if a key exists in a set or not. 
Obviously, there exists subtle but important differences. 
For example, a \bloomfilter can have false positives but not false negatives. 
However, as we will show in this paper, it is possible to address these differences through novel learning techniques and/or simple auxiliary data structures. 

In terms of performance, we observe that every CPU already has powerful SIMD capabilities and we speculate that many laptops and mobile phones will soon have a Graphics Processing Unit (GPU) or  Tensor Processing Unit (TPU).
It is also reasonable to speculate that CPU-SIMD/GPU/TPUs will be increasingly powerful as it is much easier to scale the restricted set of (parallel) math operations used by neural nets than a general purpose instruction set. 
As a result the high cost to execute a neural net or other ML models might actually be negligible in the future. 
For instance, both Nvidia and Google's TPUs are already able to perform thousands if not tens of thousands of neural net operations in a single cycle~\cite{googletpu}.
Furthermore, it was stated that GPUs will improve $1000\times$ in performance by 2025, whereas Moore's law for CPUs is essentially dead~\cite{MooresLawDead}.
By replacing branch-heavy index structures with neural networks, databases and other systems can benefit from these hardware trends. 
While we see the future of learned index structures on specialized hardware, like TPUs, this paper focuses entirely on CPUs and surprisingly shows that we can achieve significant advantages even in this case. 

It is important to note that we do not argue to completely replace traditional index structures with learned indexes.
Rather, {\bf the main contribution of this paper is to outline and  evaluate the potential of a novel approach to build indexes, which complements existing work and, arguably, opens up an entirely new research direction for a decades-old field.}
This is based on the  key observation that {\bf many  data structures can be decomposed into a learned model and an auxiliary structure} to provide the same semantic guarantees. 
The potential power of this approach comes from the fact that {\bf continuous functions, describing the data distribution, can be used to build more efficient data structures or algorithms}.
We empirically get very promising results when evaluating our approach on synthetic and real-world datasets for read-only analytical workloads.  
However, many open challenges still remain, such as how to handle write-heavy workloads, and we outline many possible directions for future work.
Furthermore, we believe that we can use the same principle to replace other components and operations commonly used in (database) systems. 
If successful, the core idea of deeply embedding learned models into algorithms and data structures could lead to a radical departure from the way systems are currently developed. 

The remainder of this paper is outlined as follows: 
In the next two sections we introduce the general idea of learned indexes using \btrees as an example. 
In Section~\ref{sec:index:hashmap} we extend this idea to \hashmaps and in Section~\ref{sec:bloomfilter} to \bloomfilters. All sections contain a separate evaluation.
Finally in Section~\ref{sec:related} we discuss related work and conclude in Section~\ref{sec:future}.

\section{Range Index}
\label{sec:btree}
\begin{figure}[t]
\begin{center}
  \includegraphics[width=0.8\linewidth]{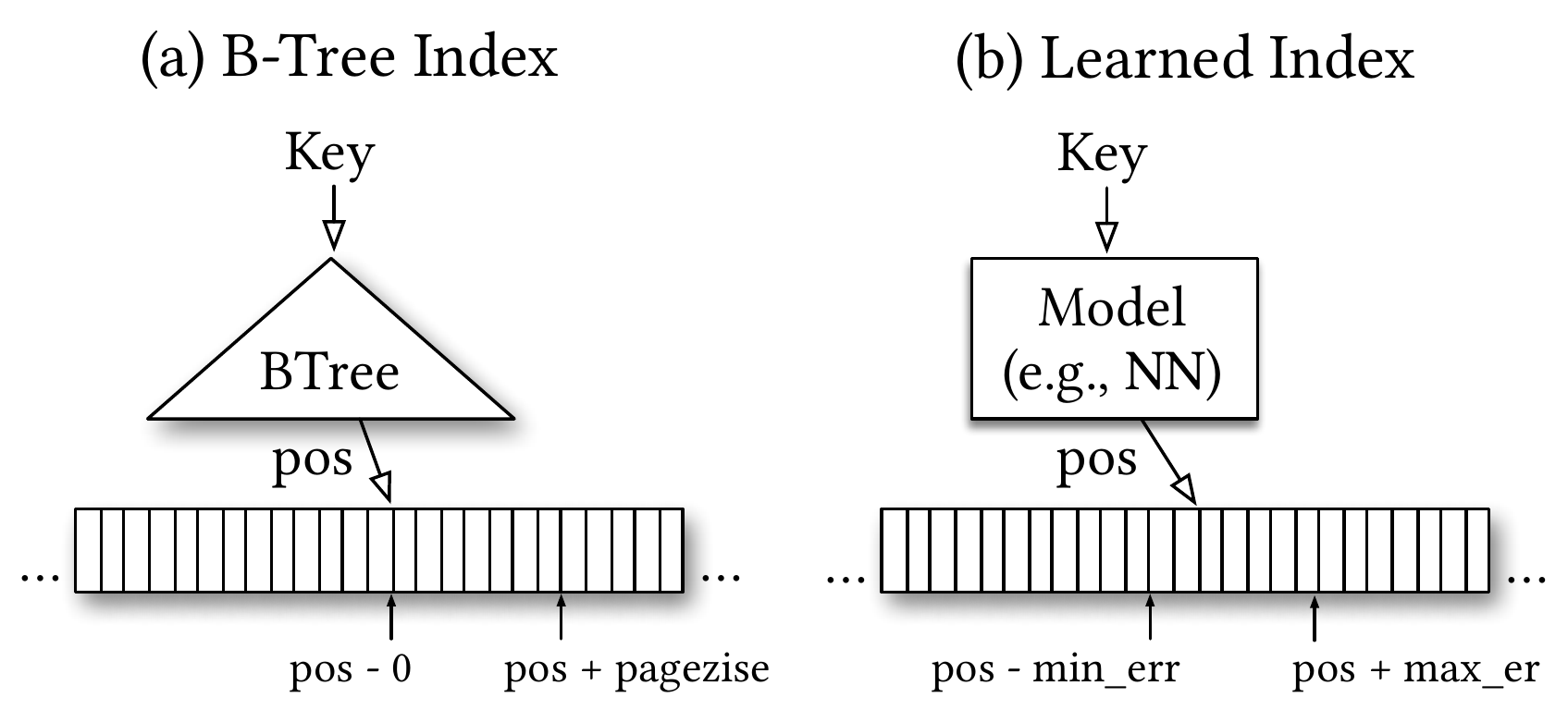}
\end{center}
\vspace{-10pt}
\caption{Why \btrees are models}
\label{fig:index}	
\end{figure}

Range index structure, like \btrees, are already models: given a key, they ``predict'' the location of a value within a key-sorted set. 
To see this, consider a \btree index in an analytics in-memory database (i.e., read-only) over the sorted primary key column as shown in Figure~\ref{fig:index}(a).
In this case, the \btree provides a mapping from a \lookup key to a position inside the sorted array of records with the guarantee that the key of the record at that position is the first key equal or higher than the \lookup key. 
The data has to be sorted to allow for efficient range requests. 
This same general concept also applies to secondary indexes where the data would be the list of \texttt{<key,record\_pointer>} pairs with the key being the indexed value and the pointer a reference to the record.\footnote{Note, that against some definitions for secondary indexes we do not consider  the \texttt{<key,record\_pointer>} pairs as part of the index; rather for secondary index the data are the \texttt{<key,record\_pointer>} pairs. 
This is similar to how indexes are implemented in key value stores \cite{SCADS,BDS3} or how B-Trees on modern hardware are designed \cite{fast}.}

For efficiency reasons it is common not to index every single key of the sorted records, rather only the  key of every n-th record, i.e., the first key of a page.
Here we only assume  fixed-length records and logical paging over a continuous memory region, i.e., a single array, not physical pages which are located in different memory regions  (physical pages and variable length records are discussed in Appendix~\ref{sec:rmi:paging}).
Indexing only the first key of every page helps to significantly reduce the number of keys the index has to store without any significant performance penalty. 
Thus, the  \btree is a model, or in ML terminology, a regression tree: 
it maps a key to a position with a min- and max-error  (a min-error of 0 and a max-error of the page-size), with a guarantee that the key can be found in that region if it exists.
Consequently, we can replace the index with other types of ML models, including neural nets, as long as 
they are also able to provide similar strong guarantees about the min- and max-error.

At first sight it may seem hard  to provide the same  guarantees with other types of ML models, but it is actually surprisingly simple.
First, the \btree only provides the strong min- and max-error guarantee over the  stored keys, not for all possible keys. 
For new data, \btrees need to be re-balanced, or in machine learning terminology re-trained, to still be able to provide the same error guarantees. 
That is, for monotonic models the only thing we need to do is to execute the model for every key and remember the worst over- and under-prediction of a position to calculate the min- and max-error.\footnote{The model has to be monotonic to also guarantee the min- and max-error for \lookup keys, which do  not exist in the stored set.}
Second, and more importantly,  the strong error bounds are not even needed. The data has to be sorted anyway to support range requests, so any error is easily corrected by a local search around the prediction (e.g., using exponential search) and thus, even allows for non-monotonic models. 
Consequently, we are able to replace \btrees with any other type of regression model, including linear regression or neural nets (see  Figure~\ref{fig:index}(b)).

Now, there are other technical challenges that we need to address before we can replace \btrees with learned indexes. For instance, \btrees have a bounded cost for inserts and \lookups and are particularly good at taking advantage of the cache.
Also, \btrees can map keys to pages which are not continuously mapped to memory or disk. 
All of these are interesting challenges/research questions and are explained in more detail, together with potential solutions, throughout this section and in the appendix.

At the same time, using other types of models as indexes can provide tremendous benefits. 
Most importantly, it has the potential to transform the $\log n$ cost of a \btree \lookup into a constant operation. 
For example, assume a dataset with 1M unique keys with a value from 1M and 2M (so the value 1,000,009 is stored at position 10). 
In this case, a simple linear model, which consists of a single multiplication and addition, can perfectly predict the position of any key for a point \lookup or range scan, whereas a \btree would require $\log n$ operations.
The beauty of machine learning, especially neural nets, is that they are able to learn a wide variety of data distributions, mixtures and other data peculiarities and patterns.
The challenge is to balance the complexity of the model with its accuracy.

For most of the discussion in this paper, we keep the simplified assumptions of this section: we only index an in-memory dense array that is sorted by key. 
This may seem restrictive, but many modern hardware optimized \btrees, e.g., FAST \cite{fast}, make exactly the same assumptions, and these indexes are quite common for in-memory database systems for their superior performance \cite{fast,ARTful} over scanning or binary search. 
However, while some of our techniques translate well to some scenarios (e.g., disk-resident data with very large blocks, for example, as used in Bigtable \cite{bigtable}), for other scenarios (fine grained paging, insert-heavy workloads, etc.) more research is needed. 
In Appendix~\ref{sec:rmi:paging} we discuss some of those challenges and potential solutions in more detail.

\subsection{What Model Complexity Can We Afford? }
\label{sec:btree:calc}
To better understand the model complexity, it is important to know how many operations can be performed in the same amount of time it takes to traverse a \btree, and what precision the model needs to achieve to be more efficient than a \btree. 

Consider a \btree that indexes 100M records with a page-size of 100.
We can think of every \btree node as a way to partition the space,  decreasing the ``error'' and narrowing the region to find the data.
We therefore say that the \btree with a page-size of 100 has a {\em precision gain} of $1/100$ per node and we need to traverse in total $log_{100} N$ nodes. 
So the first node partitions the space from $100M$ to $100M/100 = 1M$, the second from $1M$ to $1M/100=10k$ and so on, until we find the record.
Now, traversing a single \btree page with binary search takes roughly $50$ cycles  and is notoriously hard to parallelize\footnote{There exist SIMD optimized index structures such as FAST \cite{fast}, but they can only transform control dependencies to memory dependencies. These are often significantly slower than multiplications with simple in-cache data dependencies and as our experiments show SIMD optimized index structures, like FAST, are not significantly faster.}.
In contrast, a modern CPU can do 8-16 SIMD operations per cycle.
Thus, a model will be faster as long as it has a better precision gain than $1/100$ per $50 * 8 = 400$ arithmetic operations.
Note that this calculation still assumes that all \btree pages are in the cache.
A single cache-miss costs 50-100 additional cycles and would thus allow for even more complex models. 

Additionally, machine learning accelerators are entirely changing the game. 
They allow to run much more complex models in the same amount of time and offload computation from the CPU.
For example,  NVIDIA's latest Tesla V100 GPU is able to achieve 120 TeraFlops  of low-precision deep learning arithmetic operations ($\approx 60,000$  operations per cycle).
Assuming that the entire learned index fits into the GPU's memory (we show in Section~\ref{sec:btree:results} that this is a very reasonable assumption), in just 30 cycles we could execute 1 million neural net operations. 
Of course, the latency for transferring the input and retrieving the result from a GPU is still significantly higher, but this problem is not insuperable given batching and/or the recent trend to more closely integrate CPU/GPU/TPUs~\cite{intelxeonphi}.
Finally, it can be expected that the capabilities and the number of floating/int operations per second of GPUs/TPUs will continue to increase, whereas the progress on increasing the performance of executing if-statements of CPUs essentially has stagnated \cite{MooresLawDead}.
Regardless of the fact that we consider GPUs/TPUs as one of the main reasons to adopt learned indexes in practice, in this paper we focus on the more limited CPUs to better study the implications of replacing and enhancing indexes through machine learning without the impact of hardware changes.

\subsection{Range Index Models are CDF Models}
\label{sec:btree:cdf}
\begin{figure}[t]
\begin{center}
  \includegraphics[width=0.6\linewidth]{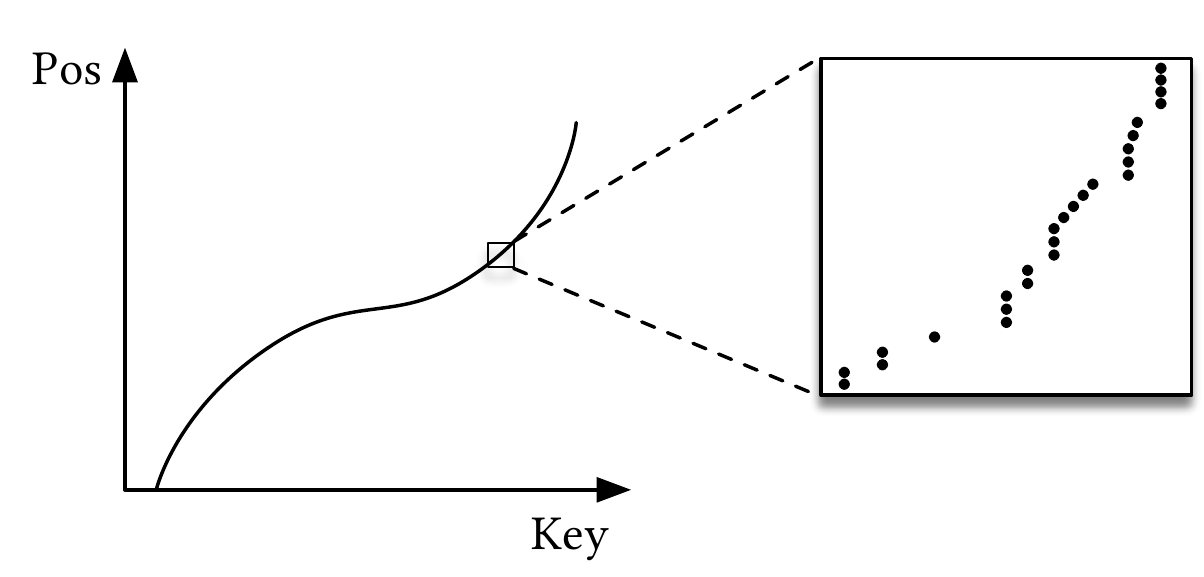}
\end{center}
\vspace{-20pt}
\caption{Indexes as CDFs}
\label{fig:cdf}	
\end{figure}

As stated in the beginning of the section, an index is a model that takes a key as an input and predicts the position of the record.
Whereas for point queries the order of the records does not matter, for range queries the data has to be sorted according to the \lookup key so that all data items in a range (e.g., in a time frame) can be efficiently retrieved. 
This leads to an interesting observation: a model that predicts the position given a key inside a sorted array effectively approximates the cumulative distribution function (CDF). 
We can model the CDF of the data to predict the position as:
\begin{equation} \label{eqn:cdf}
p = F(\text{Key}) * N
\end{equation}
where $p$ is the position estimate, $F(\text{Key})$ is the estimated cumulative distribution function for the data to estimate the likelihood to observe a key smaller or equal to the \lookup key $P(X \le \text{Key})$, and $N$ is the total number of keys  (see also Figure~\ref{fig:cdf}). 
This observation opens up a whole new set of interesting directions: 
First, it implies that indexing literally requires learning a data distribution. 
A \btree ``learns'' the data distribution by building a regression tree. 
A linear regression model would learn the data distribution by minimizing the (squared) error of a linear function. 
Second, estimating the distribution for a dataset is a well known problem and learned indexes can benefit from decades of research. 
Third, learning the CDF plays also a key role in optimizing other types of index structures and potential algorithms as we will outline later in this paper.   
Fourth, there is a long history of research on how closely theoretical CDFs approximate empirical CDFs that gives a foothold to theoretically understand the benefits of this approach \cite{dvoretzky1956asymptotic}.  We give a high-level theoretical analysis of how well our approach scales in Appendix \ref{sec:btree:theory}.

\subsection{A First, \Naive{} Learned Index }
\label{sec:btree:naive}
To better understand the requirements to replace \btrees through learned models, we used 200M web-server log records with the goal of building a secondary index over the timestamps using Tensorflow \cite{abadi2016tensorflow}. 
We trained a two-layer fully-connected neural network with 32 neurons per layer using ReLU activation functions; the timestamps are the input features and the positions in the sorted array are the labels. 
Afterwards we measured the \lookup time for a randomly selected key (averaged over several runs disregarding the first numbers) with Tensorflow and Python as the front-end. 

In this setting we achieved $\approx 1250$ predictions per second, i.e., it takes $\approx 80,000$ nano-seconds (ns) to execute the model with Tensorflow, without the search time (the time to find the actual record from the predicted position). 
As a comparison point, a \btree traversal over the same data takes $\approx 300ns$ and binary search over the entire data roughly $\approx 900ns$.
With a closer look, we find our \naive{} approach is limited in a few key ways:
\begin{enumerate}
\item
Tensorflow was designed to efficiently run larger models, not small models, and thus, has a significant invocation overhead, especially with Python as the front-end.

\item \btrees, or decision trees in general, are really good in overfitting the data with a few operations as they recursively divide the space using simple if-statements. 
In contrast, other models can be significantly more efficient to approximate the general shape of a CDF, but have problems being accurate at the individual data instance level. 
To see this, consider again Figure~\ref{fig:cdf}. 
The figure demonstrates, that from a top-level view, the CDF function appears very smooth and regular. 
However, if one zooms in to the individual records, more and more irregularities show; a well known statistical effect.
Thus models like neural nets, polynomial regression, etc. might be more CPU and space efficient to narrow down the position for an item from the entire dataset to a region of thousands, but a single neural net usually requires significantly more space and CPU time for the ``last mile'' to reduce the error further down from thousands to hundreds. 

\item  \btrees are extremely cache- and operation-efficient as they keep the top nodes always in cache and access other pages if needed. 
In contrast, standard neural nets require all weights to compute a prediction, which has a high cost in the number of multiplications.
\end{enumerate}

\section{The RM-Index}
\label{sec:rmi}
In order to overcome the challenges and explore the potential of models as index replacements or optimizations, we developed the  learning index framework (LIF), recursive-model indexes (RMI), and standard-error-based search strategies. 
We primarily focus on simple, fully-connected neural nets because of their simplicity and flexibility, but 
we believe other types of models may provide additional benefits.

\subsection{The Learning Index Framework (LIF)}
\label{sec:btree:lif}
The \framework{} can be regarded as an index synthesis system; 
given an index specification, \framework{} generates different index configurations, optimizes them, and tests them automatically. 
While \framework{} can learn simple models on-the-fly (e.g., linear regression models), it relies on Tensorflow for more complex models (e.g., NN). 
However, it never uses Tensorflow at inference.
Rather, given a  trained Tensorflow model, \framework{} automatically extracts all weights from the  model and generates efficient index structures in C++ based on the model specification. 
Our code-generation is particularly designed for small models and removes all unnecessary overhead and instrumentation that Tensorflow has to manage the larger models.
Here we leverage ideas from \cite{udfcompilation}, which already showed how to avoid unnecessary overhead from the Spark-runtime. 
As a result, we are able to execute simple models on the order of 30 nano-seconds. 
However, it should be pointed out that LIF is still an experimental framework and is instrumentalized to quickly evaluate different index configurations (e.g., ML models, page-sizes, search strategies, etc.), which introduces additional overhead in form of additional counters, virtual function calls, etc.
Also besides the vectorization done by the compiler, we do not make use of special SIMD intrinisics. 
While these inefficiencies do not matter in our evaluation as we ensure a fair comparison by always using our framework, for a production setting or when comparing the reported performance numbers with other implementations, these inefficiencies should be taking into account/be avoided.  

\subsection{The Recursive Model Index}
\label{sec:btree:staging}
As outlined in Section~\ref{sec:btree:naive} one of the key challenges of building alternative learned models to replace \btrees is the accuracy for last-mile search. 
For example, reducing the prediction error to the order of hundreds from 100M records using a single model is often difficult. 
At the same time, reducing the error to 10k from 100M, e.g., a precision gain of $100 * 100 = 10000$ to replace the first 2 layers of a \btree through a model, is much easier to achieve even with simple models. 
Similarly, reducing the error from 10k to 100 is a simpler problem as the model can focus only on a subset of the data.

\begin{figure}[t]
	\centering
    \includegraphics[width=0.7\linewidth]{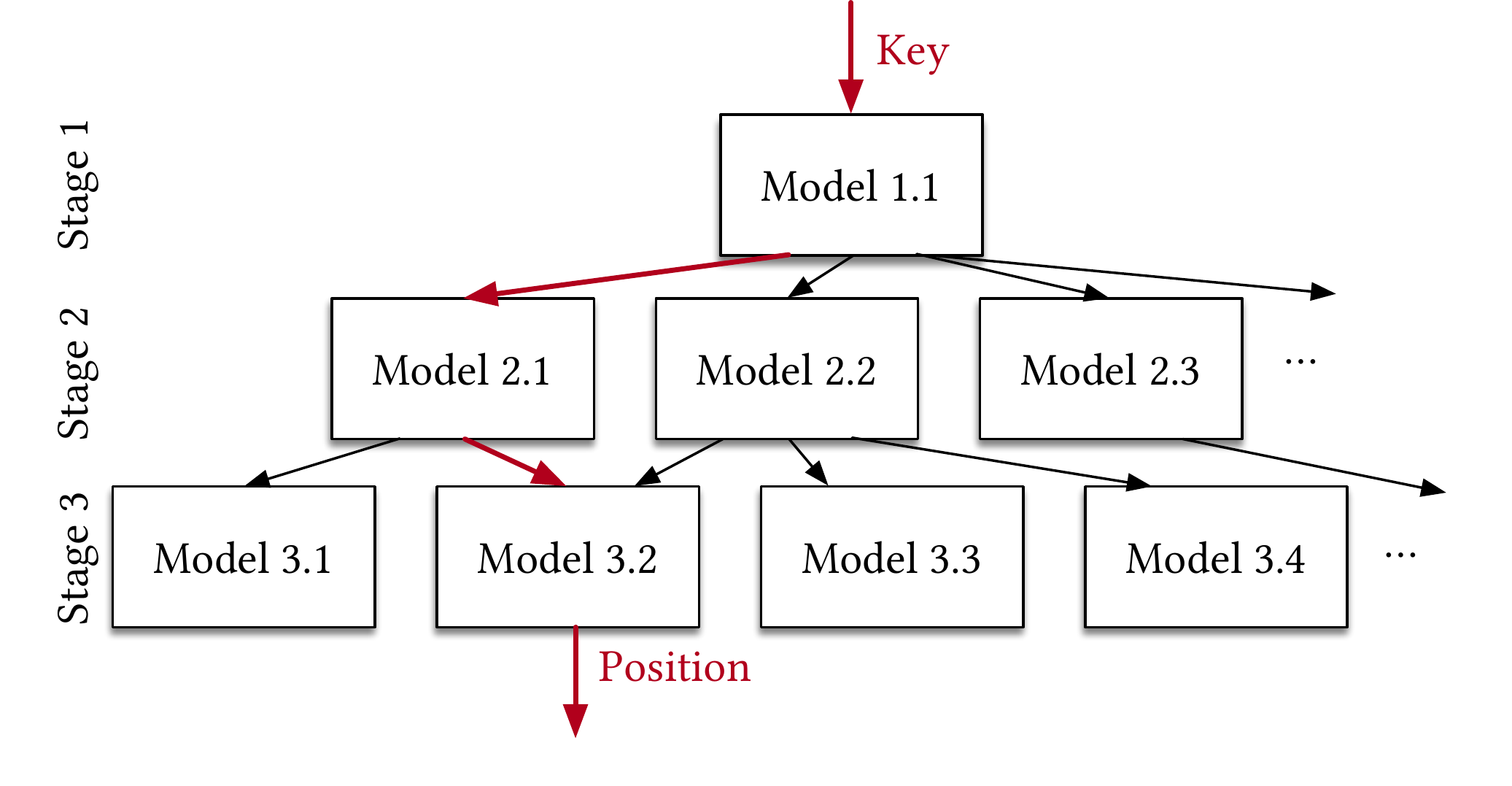}
	\caption{Staged models}
     \label{fig:staged_model}
\end{figure}

Based on that observation and inspired by the mixture of experts work \cite{moe}, we propose the recursive regression model (see Figure~\ref{fig:staged_model}). 
That is, we build a hierarchy of models, where at each stage the model takes the key as an input and based on it picks another model, until the final stage predicts the position. 
More formally, for our model $f(x)$ where $x$ is the key and $y \in [0,N)$ the position, we assume at stage $\ell$ there are $M_\ell$ models.
We train the model at stage 0, $f_0(x) \approx y$.  As such, model $k$ in stage $\ell$, denoted by $f_\ell^{(k)}$, is trained with loss: 

\begin{small}
\begin{align*}
    L_\ell  &= \sum_{(x,y)} (f_\ell^{(\lfloor M_\ell f_{\ell-1}(x)/N \rfloor)}(x) - y)^2
    & L_0 &= \sum_{(x,y)} (f_0(x) - y)^2
\end{align*}
\end{small}

Note, we use here the notation of $f_{\ell-1}(x)$ recursively executing $f_{\ell-1}(x) = f_{\ell-1}^{\left(\lfloor M_{\ell-1}f_{\ell-2}(x)/N\rfloor \right)}(x)$.
In total, we iteratively train each stage with loss $L_\ell$ to build the complete model.  
One way to think about the different models is that each model makes a prediction with a certain error about the position for the {\em key} and that the prediction is used to select the next model, which is responsible for a certain area of the key-space to make a better prediction with a lower error. 
However,  recursive model indexes do \emph{not} have to be trees.
As shown in Figure~\ref{fig:staged_model} it is possible that different models of one stage pick the same models at the stage below. 
Furthermore, each model does not necessarily cover the same amount of records like \btrees do (i.e., a \btree with a page-size of 100 covers 100 or less records).\footnote{Note, that we currently train stage-wise and not fully end-to-end. End-to-end training would be even better and remains future work.}
Finally, depending on the used models the predictions between the different stages can not necessarily be interpreted as positions estimates, rather should be considered as picking an expert which has a better knowledge about certain keys (see also \cite{moe}). 

This model architecture has several benefits:
(1) It separates model size and complexity from execution cost.
(2) It leverages the fact that it is easy to learn the overall shape of the data distribution.  
(3) It effectively divides the space into smaller sub-ranges, like a \btree, to make it easier to achieve the required ``last mile'' accuracy with fewer operations. 
(4) There is no search process required in-between the stages. 
For example, the output of {\em Model 1.1} is directly used to pick the model in the next stage. 
This not only reduces the number of instructions to manage the structure, but also allows representing the entire index as a sparse matrix-multiplication for a TPU/GPU.

\subsection{Hybrid Indexes}
\label{sec:btree:hybrid}
Another advantage of the recursive model index is, that we are able to build mixtures of models. 
For example, whereas on the top-layer a small ReLU neural net might be the best choice as they are usually able to learn a wide-range of complex data distributions, the models at the bottom of the model hierarchy might be thousands of simple linear regression models as they are inexpensive in space and execution time.
Furthermore, we can even use traditional \btrees at the bottom stage if the data is particularly hard to learn.

For this paper, we focus on 2 types of models, simple neural nets with zero to two fully-connected hidden layers and ReLU activation functions and a layer width of up to 32 neurons and \btrees (a.k.a. decision trees). 
Note, that a zero hidden-layer NN is equivalent to linear regression. 
Given an index configuration, which specifies the number of stages and the number of models per stage as an array of sizes, the end-to-end training for hybrid indexes is done as shown in Algorithm~\ref{alg:hybrid}

\begin{algorithm}
\footnotesize
\LinesNumbered
\AlgoDisplayBlockMarkers\SetAlgoBlockMarkers{}{}%
\SetAlgoNoEnd
    \SetAlgoLined
    \KwIn{int threshold, int stages[], NN\_complexity}
    \KwData{record data[], Model index[][]}
    \KwResult{trained index }
    $M$ = stages.size\;
    tmp\_records[][]\;
    tmp\_records[1][1] = all\_data\;
    \For{$i\leftarrow 1$ \KwTo $M$}{
       \For{$j\leftarrow 1$ \KwTo $stages[i]$}{
            index[i][j] = new NN trained on tmp\_records[$i$][$j$]\;
            \If{$i < M$}{
                \For{$r\in$ tmp\_records[$i$][$j$]}{
                    $p$ = index[i][j]$(r.key)$ / stages[$i+1$]\;
                    tmp\_records[$i+1$][$p$].add($r$)\;
                }
            }
        }
    }
    \For{$j\leftarrow 1$ \KwTo $index[M].size$}{
        index[$M$][$j$].calc\_err(tmp\_records[$M$][$j$])\;
        \If{$index[M][j].max\_abs\_err > threshold$}{
            index[$M$][$j$] = new \btree trained on tmp\_records[$M$][$j$]\;
        }
    }
    \Return index\;
\caption{Hybrid End-To-End Training}
\label{alg:hybrid}

\end{algorithm}

Starting from the entire dataset (line 3), it trains first the top-node model. 
Based on the prediction of this top-node model, it then picks the model from the next stage (lines 9 and 10) and adds all keys which fall into that model (line 10).
Finally, in the case of hybrid indexes, the index is optimized by replacing NN models with \btrees if  absolute min-/max-error is above a predefined threshold (lines 11-14).

Note, that we store the standard and min- and max-error for every model on the last stage. 
That has the advantage, that we can individually restrict the search space based on the used model for every key. 
Currently, we tune the various parameters of the model (i.e., number of stages, hidden layers per model, etc.) with a simple simple grid-search.
However, many potential optimizations exists to speed up the training process from ML auto tuning to sampling. 

{\bf Note, that hybrid indexes allow us to bound the worst case performance of learned indexes to the performance of \btrees.}
That is, in the case of an extremely difficult to learn data distribution, all models would be automatically replaced by \btrees, making it virtually an entire \btree. 

\subsection{Search Strategies and Monotonicity}
\label{sec:btree:search}
Range indexes usually implement an $upper\_bound(key)$ [$lower\_$ $bound(key)$] interface to find the position of the first key within the sorted array that is equal or higher [lower] than the \lookup key to efficiently support range requests.
For learned range indexes we therefore have to find the first key higher [lower] from the \lookup key based on the prediction. 
Despite many efforts, it was repeatedly reported \cite{neumannbinary} that binary search or scanning for records with small payloads are usually the fastest strategies to find a key within a sorted array as the additional complexity of alternative techniques rarely pays off.  
However, learned indexes might have an advantage here: the models actually predict the position of the key, not just the region (i.e., page) of the key.
Here we discuss two simple search strategies which take advantage of this information:

{\bf Model Biased Search:} Our default search strategy, which only varies from traditional binary search in that the first {\em middle} point is set to the value predicted by the model.

{\bf Biased Quaternary Search:} 
Quaternary search takes instead of one split point three points with the hope that the hardware pre-fetches all three data points at once to achieve better performance if the data is not in cache. 
In our implementation, we defined the initial three middle points of quaternary search as $pos - \sigma, pos, pos + \sigma$. 
That is we make a guess that most of our predictions are accurate and focus our attention first around the position estimate and then we continue with traditional quaternary search. 

For all our experiments we used the min- and max-error as the search area for all techniques. 
That is, we executed the RMI model for every key and stored the worst over- and under-prediction per last-stage model. 
While this technique guarantees to find all existing keys, for non-existing keys it might return the wrong upper or lower bound if the RMI model is not monotonic. 
To overcome this problem, one option is to force our RMI model to be monotonic, as has been studied in machine learning \cite{gupta2016monotonic,you2017deep}.

Alternatively, for non-monotonic models we can  automatically adjust the search area.
That is, if the found upper (lower) bound key is on the boundary of the search area defined by the min- and max-error, we incrementally adjust the search area. 
Yet, another possibility is, to use exponential search techniques.
Assuming a normal distributed error, those techniques on average should work as good as alternative search strategies while not requiring to store any min- and max-errors.

\subsection{Indexing Strings}
\label{sec:btree:strings}
We have primarily focused on indexing real valued keys, but many databases rely on indexing strings, and luckily, significant machine learning research has focused on modeling strings.  As before, we need to design a model of strings that is efficient yet expressive.  Doing this well for strings opens a number of unique challenges.

The first design consideration is how to turn strings into features for the model, typically called tokenization.  For simplicity and efficiency, we consider an $n$-length string to be a feature vector $\mathbf{x} \in \mathbb{R}^n$ where $\mathbf{x}_i$ is the ASCII decimal value (or Unicode decimal value depending on the strings).  Further, most ML models operate more efficiently if all inputs are of equal size.  As such, we will set a maximum input length $N$.  Because the data is sorted lexicographically, we will truncate the keys to length $N$ before tokenization.  For strings with length $n < N$, we set $\mathbf{x}_i = 0$ for $i > n$.

For efficiency, we generally follow a similar modeling approach as we did for real valued inputs.  We learn a hierarchy of relatively small feed-forward neural networks.  The one difference is that the input is not a single real value $x$ but a vector $\mathbf{x}$.  Linear models $\mathbf{w} \cdot \mathbf{x} + \mathbf{b}$ scale the number of multiplications and additions linearly with the input length $N$.  Feed-forward neural networks with even a single hidden layer of width $h$ will scale $O(hN)$ multiplications and additions.

Ultimately, we believe there is significant future research that can optimize learned indexes for string keys. 
For example, we could easily imagine other tokenization algorithms.  There is a large body of research in natural language processing on string tokenization to break strings into more useful segments for ML models, e.g., wordpieces in translation \cite{wu2016google}.  
Further, it might be interesting to combine the idea of suffix-trees with learned indexes as well as explore more complex model architectures  (e.g., recurrent and convolutional neural networks).

\subsection{Training}
\label{sec:btree:training}
While the training (i.e., loading) time is not the focus of this paper, it should be pointed out that all of our models, shallow NNs or even simple linear/multi-variate regression models, train relatively fast.
Whereas simple NNs can be efficiently trained using stochastic gradient descent and can converge in less than one to a few passes over the randomized data, a closed form solution exists for linear multi-variate models (e.g., also 0-layer NN) and they can be trained in a single pass over the sorted data. 
Therefore, for 200M records training a simple RMI index does not take much longer than a few seconds, (of course, depending on how much auto-tuning is performed); neural nets can train on the order of minutes per model, depending on the complexity.
Also note that training the top model over the entire data is usually not necessary as those models converge often even before a single scan over the entire randomized data.  This is in part because we use simple models and do not care much about the last few digit points in precision, as it has little effect on indexing performance. 
Finally, research on improving learning time from the ML community \cite{duchi2011adaptive,you2017imagenet} applies in our context and we expect a lot of future research in this direction.

\begin{figure*}[t]
\begin{center}
  \includegraphics[width=\textwidth]{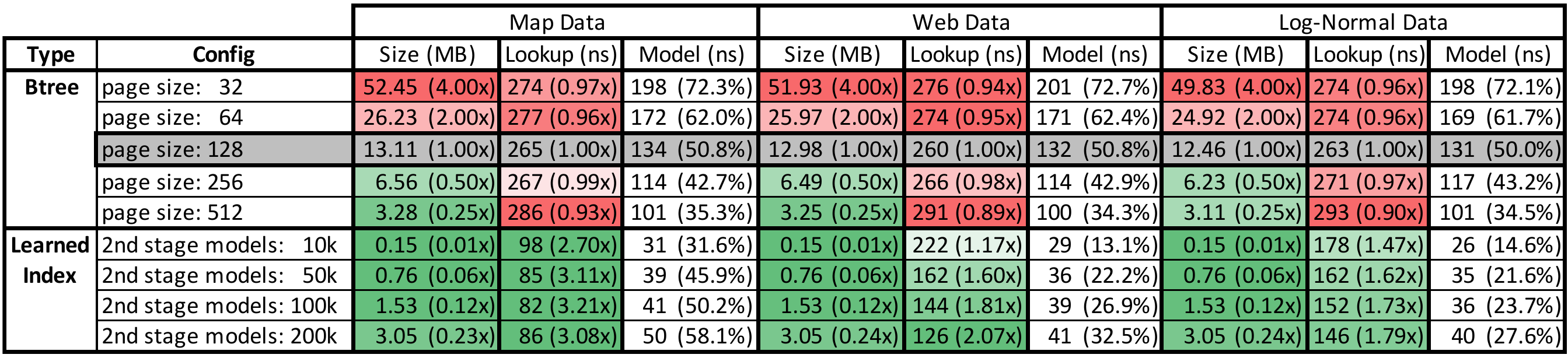}
\end{center}
\vspace{-15pt}
\caption{Learned Index vs \btree }
\label{fig:btree_int_cr_result}	
\end{figure*}

\subsection{Results}
\label{sec:btree:results}
We evaluated learned range indexes in regard to their space and speed on several real and synthetic data sets against other read-optimized index structures.

\subsubsection{Integer Datasets}
As a first experiment we compared learned indexes using a 2-stage RMI model and different second-stage sizes (10k, 50k, 100k, and 200k) with a read-optimized \btree  with different page sizes on three different integer data sets. 
For the data we used 2 real-world datasets, (1) Weblogs and (2) Maps~\cite{OSMdata}, and (3) a synthetic dataset, Lognormal. 
The Weblogs dataset contains $200$M log entries for every request to a major university web-site over several years. We use the unique request timestamps as the index keys. 
This dataset is almost a worst-case scenario for the learned index as it contains very complex time patterns caused by class schedules, weekends, holidays, lunch-breaks, department events, semester breaks, etc., which are notoriously hard to learn.
For the maps dataset we indexed the longitude of $\approx 200$M user-maintained features (e.g., roads, museums, coffee shops) across the world.
Unsurprisingly, the longitude of locations is relatively linear and has fewer irregularities than the Weblogs dataset. 
Finally, to test how the index works on heavy-tail distributions, we generated a synthetic dataset of 190M unique values sampled from a log-normal distribution with $\mu=0$ and $\sigma=2$.  The values are scaled up to be integers up to 1B.  This data is of course highly non-linear, making the CDF more difficult to learn using neural nets.
For all B-Tree experiments we used 64-bit keys and 64-bit payload/value.

As our baseline, we used a production quality \btree implementation which is similar to the stx::btree but with further cache-line optimization, dense pages (i.e., fill factor of $100\%$), and very competitive performance.
To tune the 2-stage learned indexes we used simple grid-search over neural nets with zero to two hidden layers and layer-width ranging from 4 to 32 nodes.
In general we found that a simple (0 hidden layers) to semi-complex (2 hidden layers and 8- or 16-wide) models for the first stage work the best. 
For the second stage, simple, linear models, had the best performance. 
This is not surprising as for the last mile it is often not worthwhile to execute complex models, and linear models can be learned optimally.

{\bf Learned Index vs \btree performance:} The main results are shown in Figure~\ref{fig:btree_int_cr_result}. 
Note, that the page size for \btrees indicates the number of keys per page not the  size in Bytes, which is actually larger.  
As the main metrics we show the size in MB, the total \lookup time in nano-seconds, and the time to execution the model (either \btree traversal or ML model) also in nano-seconds and as a percentage compared to the total time in paranthesis. 
Furthermore, we show the speedup and space savings compared to a \btree with page size of 128 in parenthesis as part of the size and lookup column.
We choose a page size of 128 as the fixed reference point as it provides the best lookup performance for \btrees (note, that it is always easy to save space at the expense of lookup performance by simply having no index at all). 
The color-encoding in the speedup and size columns indicates how much faster or slower (larger or smaller) the index is against the reference point. 

As can be seen, the learned index dominates the \btree index in almost all configurations by being up to $1.5 - 3\times$ faster while being up to two orders-of-magnitude smaller. 
Of course, \btrees can be further compressed at the cost of CPU-time for decompressing. 
However, most of these optimizations are orthogonal and apply equally (if not more) to neural nets.
For example, neural nets can be compressed by using 4- or 8-bit integers instead of 32- or 64-bit floating point values to represent the model parameters (a process referred to as quantization). This level of compression can unlock additional gains for learned indexes. 

Unsurprisingly the second stage size has a significant impact on the index size and \lookup performance.
Using 10,000 or more models in the second stage is particularly impressive with respect to the analysis in \S\ref{sec:btree:calc}, as it demonstrates that our first-stage model can make a much larger jump in precision than a single node in the \btree.
Finally, we do not report on hybrid models or other search techniques than binary search for these datasets as they did not provide significant benefit.

{\bf Learned Index vs Alternative Baselines:} 
In addition to the detailed evaluation of learned indexes against our read-optimized \btrees, we also compared learned indexes against other alternative baselines, including third party implementations, under fair conditions. 
In the following, we discuss some alternative baselines and compare them against learned indexes if appropriate:

{\underline{Histogram}:} \btrees approximate the CDF of the underlying data distribution. 
An obvious question is whether histograms can be used as a CDF model. 
In principle the answer is yes, but to enable fast data access, the histogram must be a low-error approximation of the CDF. Typically this requires a large number of buckets, which makes it expensive to search the histogram itself.
This is especially true, if the buckets have varying bucket boundaries to efficiently handle data skew, so that only few buckets are empty or too full.
The obvious solutions to this issues would yield a \btree, and histograms are therefore not further discussed.

{\underline{ Lookup-Table}:} 
A simple alternative to \btrees are (hierarchical) lookup-tables. 
Often lookup-tables have a fixed size and structure (e.g., 64 slots for which each slot points to another 64 slots, etc.).
The advantage of lookup-tables is that because of their fixed size they can be highly optimized using AVX instructions.
We included a comparison against a 3-stage lookup table, which is constructed by taking every 64th key and putting it into an array including padding to make it a multiple of 64. 
Then we repeat that process one more time over the array without padding, creating two arrays in total. 
To lookup a key, we use binary search on the top table followed by an AVX optimized branch-free scan \cite{avxscan} for the second table and the data itself. 
This configuration leads to the fastest lookup times compared to alternatives (e.g., using scanning on the top layer, or binary search on the 2nd array or the data).

{\underline{ FAST}:} 
FAST \cite{fast} is a highly SIMD optimized data structure. We used the code from \cite{fastneumann} for the comparison. However, it should be noted that FAST always requires to allocate memory in the power of 2 to use the branch free SIMD instructions, which can lead to significantly larger indexes. 

{\underline{ Fixed-size \btree \& interpolation search}:} Finally, as proposed in a recent blog post \cite{neumannblog} we created a fixed-height \btree with interpolation search. The \btree height is set, so that the total size of the tree is 1.5MB, similar to our learned model.

{\underline{ Learned indexes without overhead}:} 
For our learned index we used a 2-staged RMI index with a multivariate linear regression model at the top and simple linear models at the bottom. 
We used simple automatic feature engineering for the top model by automatically creating and selecting features in the form of $key$, $log(key)$, $key^2$, etc.
Multivariate linear regression is an interesting alternative to NN as it is particularly well suited to fit nonlinear patterns with only a few operations. 
Furthermore, we implemented the learned index outside of our benchmarking framework to ensure a fair comparison. 

\begin{figure}[t]
\begin{center}
  \includegraphics[width=0.6\linewidth]{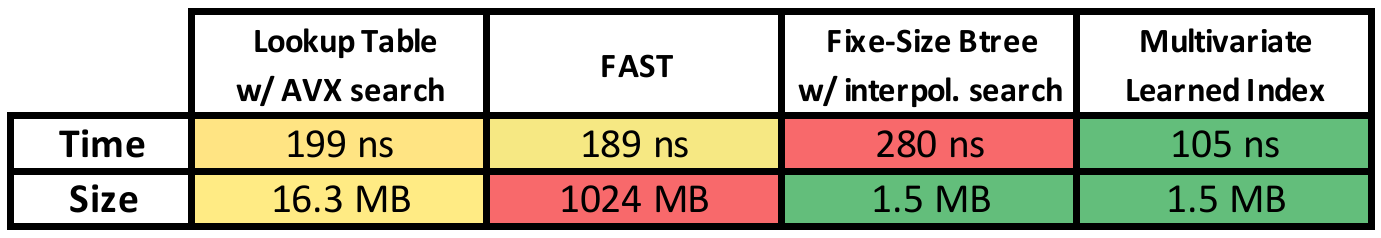}
\end{center}
\vspace{-15pt}
\caption{Alternative Baselines }
\label{fig:alt_baselines}	
\end{figure}

For the comparison we used  the Lognormal data with a payload of an eight-byte pointer. 
The results can be seen in Figure~\ref{fig:alt_baselines}. 
As can be seen for the dataset under fair conditions, learned indexes provide the best overall performance while saving significant amount of memory. It should be noted, that the FAST index is big because of the alignment requirement. 

While the results are very promising, we by no means claim that learned indexes will always be the best choice in terms of size or speed. 
Rather, learned indexes provide a new way to think about indexing and much more research is needed to fully understand the implications.

\subsubsection{String Datasets}
\begin{figure}[t]
\begin{center}
  \includegraphics[width=0.6\linewidth]{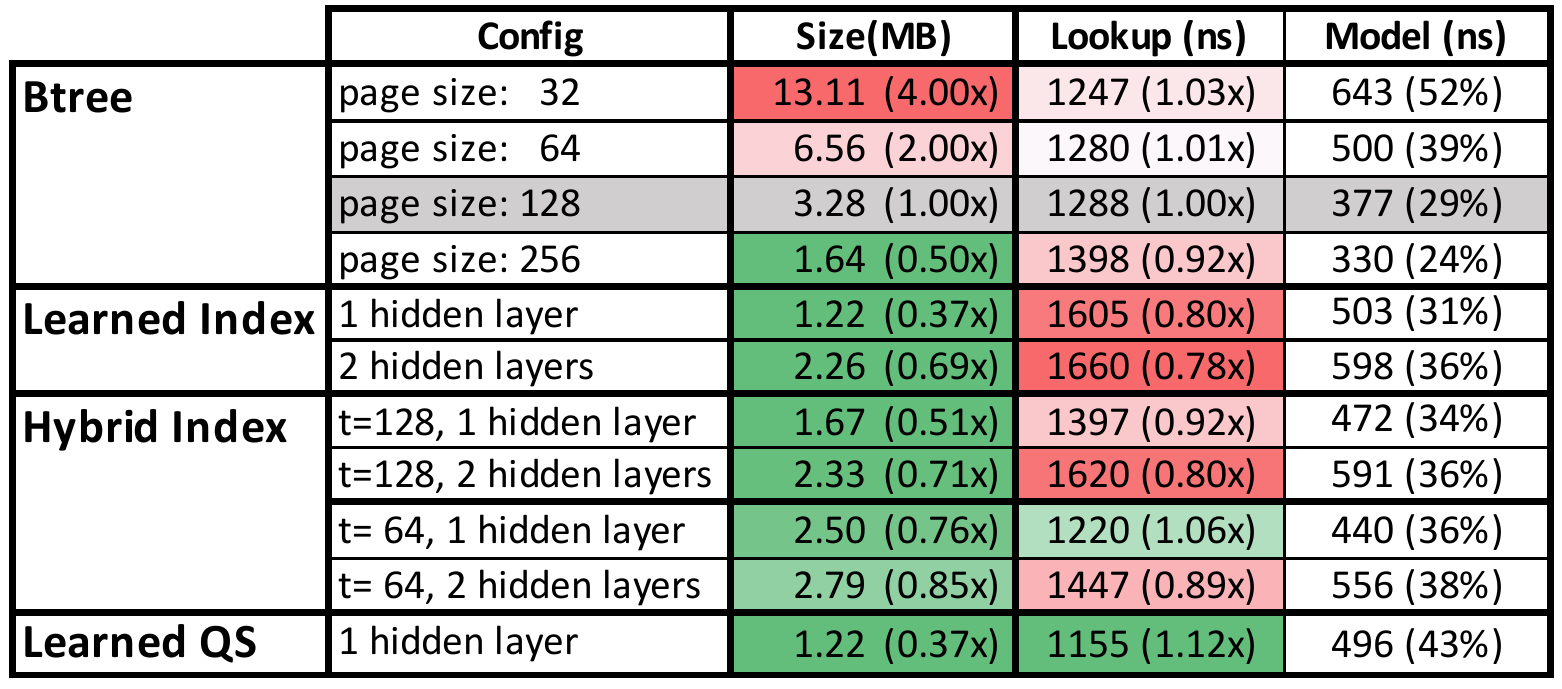}
\end{center}
\vspace{-12pt}
\caption{String data: Learned Index vs \btree }
\label{fig:string_result}	
\end{figure}

We also created a secondary index over 10M non-continuous document-ids of a large web index used as part of a real product at Google to test how learned indexes perform on strings.
The results for the string-based document-id dataset are shown in Figure~\ref{fig:string_result}, which also now includes hybrid models. 
In addition, we include our best model in the table, which is a non-hybrid RMI model index  with quaternary search, named ``Learned QS'' (bottom of the table).  
All RMI indexes used 10,000 models on the 2nd stage and for hybrid indexes we used two thresholds, 128 and 64, as the maximum tolerated absolute error for a model before it is replaced with a \btree. 

As can be seen, the speedups for learned indexes over \btrees for strings are not as prominent. Part of the reason is the comparably high cost of model execution, a problem that GPU/TPUs would remove. 
Furthermore, searching over strings is much more expensive thus higher precision often pays off;  the reason why hybrid indexes, which replace bad performing models through \btrees,  help to improve performance. 

Because of the cost of searching, the different search strategies make a bigger difference.
For example, the search time for a NN with 1-hidden layer and biased binary search is $1102ns$ as shown in Figure~\ref{fig:string_result}. 
In contrast, our biased quaternary search with the same model only takes $658ns$, a significant improvement. 
The reason why biased search and quaternary search perform better is that they take the model error into account.

\section{Point Index}
\label{sec:index:hashmap}
\begin{figure}[t]
	\centering
    \includegraphics[width=0.7\linewidth]{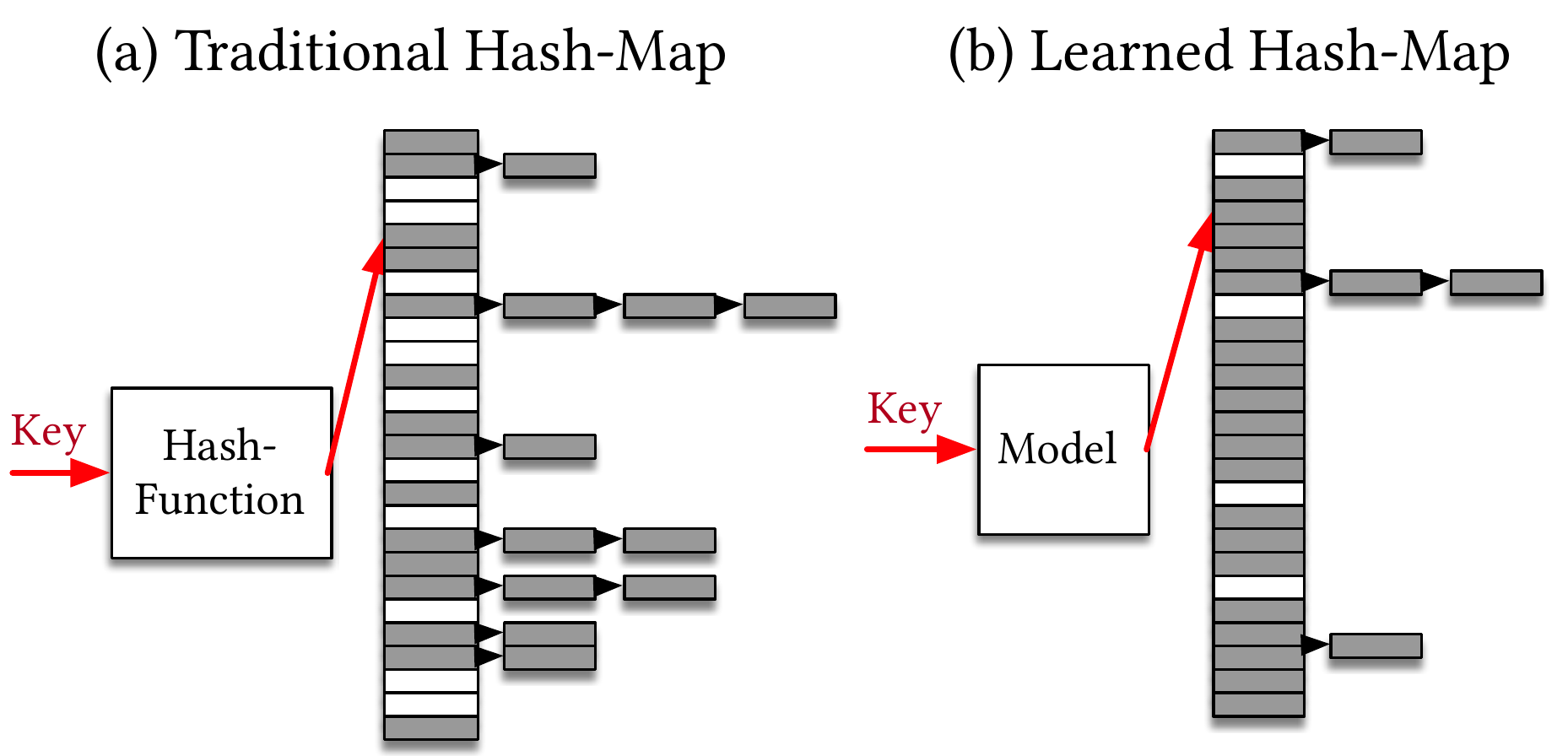}

	\caption{Traditional \hashmap vs Learned \hashmap }
     \label{fig:hash_map}
     \vspace*{-5pt}
\end{figure}

Next to range indexes, \hashmaps for point \lookups play a similarly important role in DBMS.
Conceptually \hashmaps use a hash-function to deterministically map keys to positions inside an array (see Figure~\ref{fig:hash_map}(a)).
The key challenge for any efficient \hashmap implementation is to prevent too many distinct keys from being mapped to the same position inside the \hashmap, henceforth referred to as a {\it conflict}.
For example, let's assume 100M records and a \hashmap size of 100M. 
For a hash-function which uniformly randomizes the keys, the number of expected conflicts can be derived similarly to the birthday paradox and in expectation would be around $33\%$ or $33$M slots. 
For each of these conflicts, the \hashmap architecture needs to deal with this conflict. 
For example, separate chaining \hashmaps would create a linked-list  to handle the conflict (see Figure~\ref{fig:hash_map}(a)). 
However, many alternatives exist including secondary probing, using buckets with several slots, up to simultaneously using more than one hash function  (e.g., as done by Cuckoo Hashing \cite{cuckoo}).

However, regardless of the \hashmap architecture, conflicts can have a significant impact of the performance and/or storage requirement, and machine learned models might provide an alternative to reduce the number of conflicts.
While the idea of learning models as a hash-function is not new, existing techniques do not take advantage of the underlying data distribution. 
For example, the various perfect hashing techniques \cite{perfecthashing} also try to avoid conflicts but the data structure used as part of the hash functions grow with the data size; a property learned models might not have (recall, the example of indexing all keys between 1 and 100M). 
To our knowledge it has not been explored if it is possible to learn models which yield more efficient point indexes.

\subsection{The Hash-Model Index}
Surprisingly, learning the CDF of the key distribution is one potential way to learn a better hash function. 
However, in contrast to range indexes, we do not aim to store the records compactly or in strictly sorted order.
Rather we can scale the CDF by the targeted size $M$ of the \hashmap and use $h(K) = F(K) * M$, with key $K$ as our hash-function. 
If the model $F$ perfectly learned the empirical CDF of the keys, no conflicts would exist.
Furthermore, the hash-function is orthogonal to the actual \hashmap architecture and can be combined with separate chaining or any other \hashmap type.

For the model, we can again leverage the recursive model architecture from the previous section. 
Obviously, like before, there exists a trade-off between the size of the index and performance, which is influenced by the model and dataset. 

Note, that how inserts, \lookups, and conflicts are handled is dependent on the  \hashmap architecture. 
As a result, the benefits learned hash functions provide over traditional hash functions, which map keys to a uniformly distributed space depend on two key factors:
(1) How accurately the model represents the observed CDF. For example, if the data is generated by a uniform distribution, a simple linear model will be able to learn the general data distribution, but the resulting hash function will not be better than any sufficiently randomized hash function. 
(2) Hash map architecture: depending on the architecture, implementation details, the payload (i.e., value), the conflict resolution policy, as well as how much more memory (i.e., slots) will or can be allocated, significantly influences the performance. For example, for  small keys and small or no values, traditional hash functions with Cuckoo hashing will probably work well, whereas larger payloads or distributed hash maps might benefit more from avoiding conflicts, and thus from learned hash functions.

\subsection{Results}
\label{sec:index:hashmap:results}

We evaluated the conflict rate of  learned hash functions over the three integer data sets from the previous section. 
As our model hash-functions we used the 2-stage RMI models from the previous section with 100k models on the 2nd stage and without any hidden layers. 
As the baseline we used a simple MurmurHash3-like hash-function and compared the number of conflicts for a table with the same number of slots as records.

\begin{figure}
	\centering
    \includegraphics[width=0.7\linewidth]{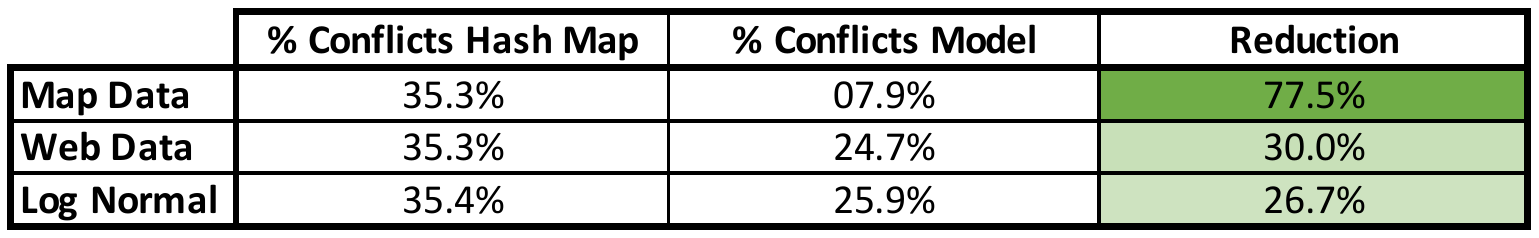}
	\caption{Reduction of Conflicts }
     \label{fig:hash_conflicts}
\end{figure}

As can be seen in Figure~\ref{fig:hash_conflicts}, the learned models can reduce the number of conflicts by up to $77\%$ over our datasets by learning the empirical CDF at a reasonable cost; the execution time is the same as the model execution time in Figure~\ref{fig:btree_int_cr_result},  around 25-40ns. 

How beneficial the reduction of conflicts is given the model execution time depends on the \hashmap architecture, payload, and many other factors.
For example, our experiments (see Appendix~\ref{appendix:chainedhash}) show that for a separate chaining \hashmap architecture with 20 Byte records learned hash functions can reduce the wasted amount of storage by up to $80\%$ at an increase of only 13ns in latency compared to random hashing. 
The reason why it only increases the latency by 13ns and not 40ns is, that often fewer conflicts also yield to fewer cache misses, and thus better performance. 
On the other hand, for very small payloads Cuckoo-hashing with standard hash-maps probably remains the best choice. 
However, as we show in Appendix~\ref{appendix:alternativehash}, for larger payloads a chained-hashmap with learned hash function can be faster than cuckoo-hashing and/or traditional randomized hashing. 
Finally, we see the biggest potential for distributed settings. 
For example, NAM-DB~\cite{nam_db} employs a hash function to \lookup data on remote machines using RDMA. 
Because of the extremely high cost for every conflict (i.e., every conflict requires an additional RDMA request which is in the order of micro-seconds), the model execution time is negligible and even small reductions in the conflict rate can significantly improve the overall performance.
To conclude, learned hash functions are independent of the used \hashmap architecture and depending on the \hashmap architecture their complexity may or may not pay off.

\section{Existence Index}
\label{sec:bloomfilter}
The last common index type of DBMS are existence indexes, most importantly \bloomfilters, a space efficient  probabilistic data structure to test whether an element is a member of a set.
They are commonly used to determine if a key exists on cold storage. 
For example, Bigtable uses them to determine if a key is contained in an SSTable \cite{bigtable}. 

\begin{figure*}[t]
	\centering
    \includegraphics[width=1\textwidth]{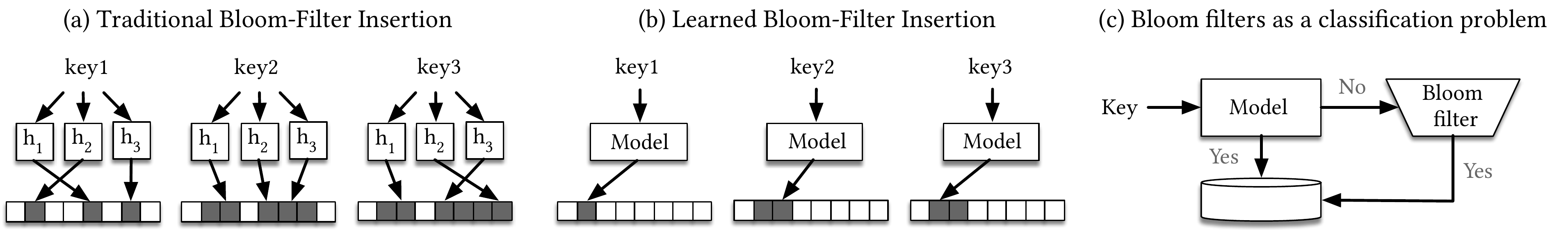}
	\caption{\bloomfilters Architectures} 
     \label{fig:learned_bloom_filter}
\end{figure*}

Internally, \bloomfilters use a bit array of size $m$ and $k$ hash functions, which each map a key to one of the $m$ array positions (see Figure\ref{fig:learned_bloom_filter}(a)).
To add an element to the set, a key is fed to the $k$ hash-functions and the bits of the returned positions are set to 1. 
To test if a key is a member of the set, the key is again fed into the $k$ hash functions to receive $k$ array positions. 
If any of the bits at those $k$ positions is 0, the key is not a member of a set. 
In other words, a \bloomfilter does guarantee that there exists {\em no false negatives}, but has potential {\em false positives}. 

While \bloomfilters are highly space-efficient, they can still occupy a significant amount of memory. 
For example for one billion records roughly $\approx 1.76$ Gigabytes are needed. 
For a FPR of $0.01\%$  we would require $\approx 2.23$ Gigabytes. 
There have been several attempts to improve the efficiency of \bloomfilters \cite{compressed_bloomfilter}, but the general observation remains.

Yet, if there is some structure to determine what is inside versus outside the set, which can be learned, it might be possible to construct more efficient representations.
Interestingly, for existence indexes for database systems, the latency and space requirements are usually quite different than what we saw before.
Given the high latency to access cold storage (e.g., disk or even band), we can afford more complex models while the main objective is to minimize the space for the index and the number of false positives.
We outline two potential ways to build existence indexes using lea
\subsection{Learned \bloomfilters}
\label{sec:bloomfilter:models}
While both range and point indexes learn the distribution of keys, existence indexes need to learn a function that separates keys from everything else.  Stated differently, a good hash function for a point index is one with few collisions among keys, whereas a good hash function for a \bloomfilter would be one that has lots of collisions among keys and lots of collisions among non-keys, but few collisions of keys and non-keys.  We consider below how to learn such a function $f$ and how to incorporate it into an existence index.

While traditional \bloomfilters guarantee a false negative rate (FNR) of zero and a specific false positive rate (FPR) for any set of queries chosen a-priori \cite{broder2004network}, we follow the notion that we want to provide a specific FPR for \emph{realistic queries} in particular while maintaining a FNR of zero.  That is, we measure the FPR over a heldout dataset of queries, as is common in evaluating ML systems \cite{fawcett2006introduction}.  While these definitions differ, we believe the assumption that we can observe the distribution of queries, e.g., from historical logs, holds in many applications, especially within databases\footnote{We would like to thank Michael Mitzenmacher for valuable conversations in articulating the relationship between these definitions as well as improving the overall chapter through his insightful comments.}.

Traditionally, existence indexes make no use of the distribution of keys nor how they differ from non-keys, but learned \bloomfilters can.   For example, if our database included all integers $x$ for $0 \leq x < n$, the existence index could be computed in constant time and with almost no memory footprint by just computing $f(x) \equiv \mathbb{1}[0 \leq x < n]$.  

In considering the data distribution for ML purposes, we must consider a dataset of non-keys.  In this work, we consider the case where non-keys come from observable historical queries and we assume that future queries come from the same distribution as historical queries.  When this assumption does not hold, one could use randomly generated keys, non-keys generated by a machine learning model \cite{goodfellow2014generative}, importance weighting to directly address covariate shift \cite{bickel2009discriminative}, or adversarial training for robustness \cite{tramer2017ensemble}; we leave this as future work.  
We denote the set of keys by $\mathcal{K}$ and the set of non-keys by $\mathcal{U}$.

\subsubsection{\bloomfilters as a Classification Problem}
\label{sec:bloomfilter:classification}
One way to frame the existence index is as a binary probabilistic classification task.  
That is, we want to learn a model $f$ that can predict if a query $x$ is a key or non-key.
For example, for strings we can train a recurrent neural network (RNN) or convolutional neural network (CNN)  \cite{sutskever2014sequence,graves2013generating} with $\mathcal{D} = \{ (x_i, y_i=1) | x_i \in \mathcal{K} \} \cup \{ (x_i, y_i = 0) | x_i \in \mathcal{U} \}$.  
Because this is a binary classification task, our neural network has a sigmoid activation to produce a probability and is trained to minimize the log loss:
$L = \sum_{(x,y) \in \mathcal{D}} y\log f(x) + (1-y)\log(1-f(x)).$

The output of $f(x)$ can be interpreted as the probability that $x$ is a key in our database.
Thus, we can turn the model into an existence index by choosing a threshold $\tau$ above which we will assume that the key exists in our database.  
Unlike \bloomfilters, our model will likely have a non-zero FPR and FNR; in fact, as the FPR goes down, the FNR will go up.  In order to preserve the no false negatives constraint of existence indexes, we create an overflow \bloomfilter.  That is, we consider $\mathcal{K}_\tau^{-} = \{ x \in \mathcal{K} | f(x) < \tau \}$ to be the set of false negatives from $f$ and create a \bloomfilter for this subset of keys.  We can then run our existence index as in Figure~\ref{fig:learned_bloom_filter}(c): if $f(x) \geq \tau$, the key is believed to exist; otherwise, check the overflow \bloomfilter.

One question is how to set $\tau$ so that our learned \bloomfilter has the desired FPR $p^*$.  We denote the FPR of our model by ${\rm FPR}_\tau \equiv \frac{\sum_{x \in \tilde{\mathcal{U}}} \mathbb{1}(f(x) > \tau)}{|\tilde{\mathcal{U}}|}$ where $\tilde{\mathcal{U}}$ is a held-out set of non-keys. We denote the FPR of our overflow \bloomfilter by ${\rm FPR}_B$. The overall FPR of our system therefore is ${\rm FPR}_O = {\rm FPR}_\tau + (1 - {\rm FPR}_\tau){\rm FPR}_B$ \cite{mitzenmacher2018model}
\footnote{We again thank Michael Mitzenmacher for identifying this relation when reviewing our paper.}.  
For simplicity, we set ${\rm FPR}_\tau = {\rm FPR}_B = \frac{p^*}{2}$ so that ${\rm FPR}_O \leq p^*$.  We tune $\tau$ to achieve this FPR on $\tilde{\mathcal{U}}$.

This setup is effective in that the learned model can be fairly small relative to the size of the data.  Further, because \bloomfilters scale with the size of key set, the overflow \bloomfilter will scale with the FNR. 
We will see experimentally that this combination is effective in decreasing the memory footprint of the existence index.  Finally, the learned model computation can benefit from machine learning accelerators, whereas traditional \bloomfilters tend to be heavily dependent on the random access latency of the memory system.

\subsubsection{\bloomfilters with Model-Hashes}
\label{sec:bloomfilter:hashes}
An alternative approach to building existence indexes is to learn a hash function with the goal to maximize collisions among keys and among non-keys while minimizing collisions of keys and non-keys.
Interestingly, we can use the same probabilistic classification model as before to achieve that. 
That is, we can create a hash function $d$, which maps $f$ to a bit array of size $m$ by scaling its output as $d = \lfloor f(x) * m \rfloor$ 
 As such, we can use $d$ as a hash function just like any other in a \bloomfilter.  This has the advantage of $f$ being trained to map most keys to the higher range of bit positions and non-keys to the lower range of bit positions (see Figure\ref{fig:learned_bloom_filter}(b)).  A more detailed explanation of the approach is given in Appendix \ref{sec:appendix:bloomfilter}.

\subsection{Results}
\label{sec:bloomfilter:results}
In order to test this idea experimentally, we explore the application of an existence index for keeping track of blacklisted phishing URLs.  We consider data from Google's transparency report as our set of keys to keep track of.  This dataset consists of 1.7M unique URLs.  We use a negative set that is a mixture of random (valid) URLs and whitelisted URLs that could be mistaken for phishing pages.
We split our negative set randomly into train, validation and test sets.
We train a character-level RNN (GRU \cite{DBLP:conf/emnlp/ChoMGBBSB14}, in particular) to predict which set a URL belongs to; we set $\tau$ based on  the validation set and also report the FPR on the test set.

\begin{figure}[t]
\begin{center}
  \includegraphics[width=0.6\linewidth]{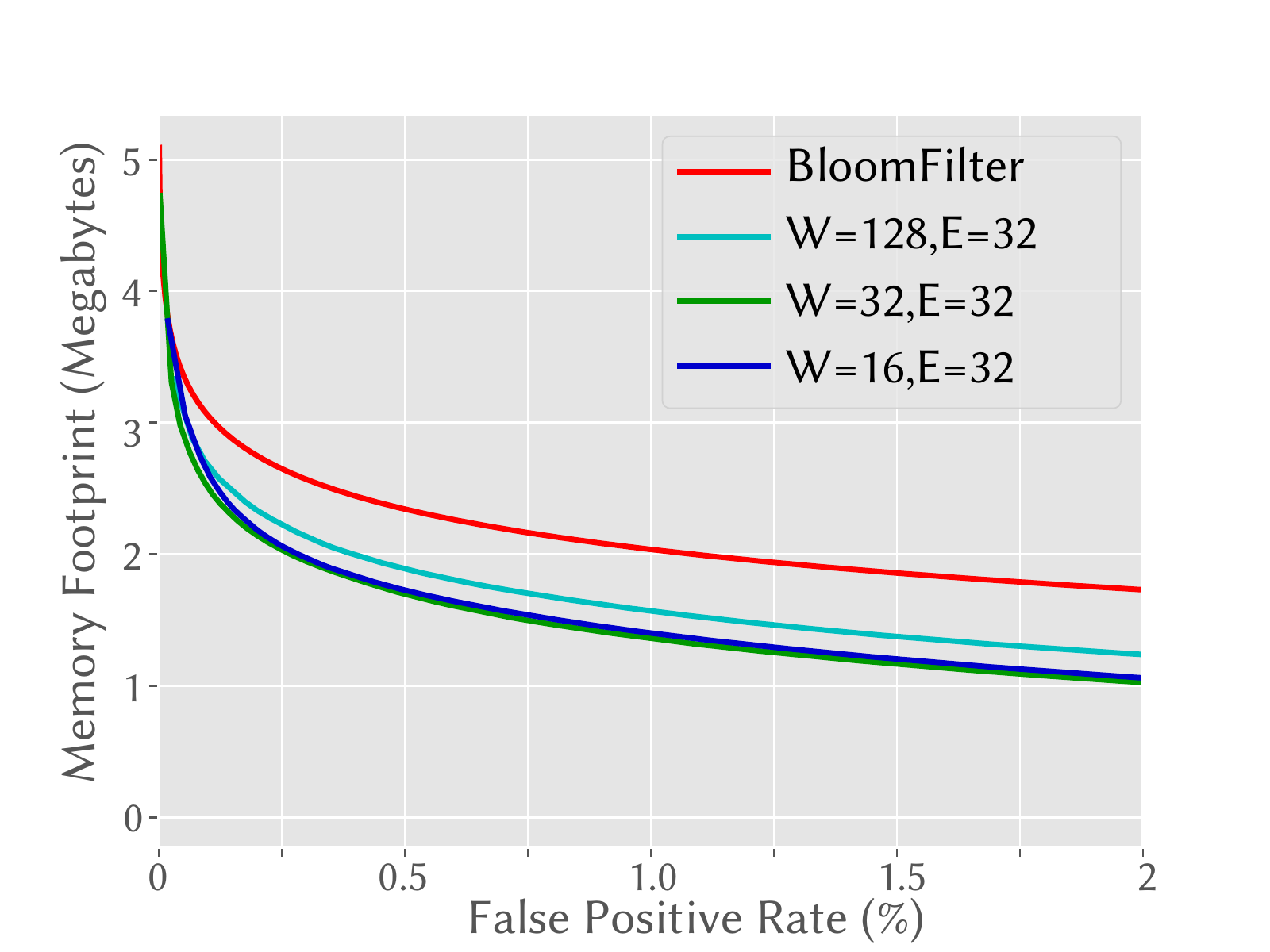}
\end{center}
\vspace{-15pt}
\caption{Learned \bloomfilter improves memory footprint at a wide range of FPRs. (Here $W$ is the RNN width and $E$ is the embedding size for each character.)}
\label{fig:bloom_results}	
\end{figure}

A normal \bloomfilter with a desired 1\% FPR requires 2.04MB. 
We consider a 16-dimensional GRU with a 32-dimen\-sional embedding for each character; this model is 0.0259MB.  When building a comparable learned index, we set $\tau$ for 0.5\% FPR on the validation set; this gives a FNR of 55\%. (The FPR on the test set is 0.4976\%, validating the chosen threshold.)  As described above, the size of our \bloomfilter scales with the FNR.  Thus, we find that our model plus the spillover \bloomfilter uses 1.31MB, a 36\% reduction in size. 
If we want to enforce an overall FPR of 0.1\%, we have a FNR of 76\%, which brings the total \bloomfilter size down from 3.06MB to 2.59MB, a 15\% reduction in memory.  We observe this general relationship in Figure \ref{fig:bloom_results}.  Interestingly, we see how different size models balance the accuracy vs. memory trade-off differently.

We consider briefly the case where there is covariate shift in our query distribution that we have not addressed in the model.  When using validation and test sets with only random URLs we find that we can save 60\% over a \bloomfilter with a FPR of 1\%.  When using validation and test sets with only the whitelisted URLs we find that we can save 21\% over a Bloom filter with a FPR of 1\%.  Ultimately, the choice of negative set is application specific and covariate shift could be more directly addressed, but these experiments are intended to give intuition for how the approach adapts to different situations.

Clearly, the more accurate our model is, the better the savings in \bloomfilter size.  One interesting property of this is that there is no reason that our model needs to use the same features as the \bloomfilter.  For example, significant research has worked on using ML to predict if a webpage is a phishing page \cite{abu2007comparison,basnet2008detection}.  Additional features like WHOIS data or IP information could be incorporated in the model, improving accuracy, decreasing \bloomfilter size, and keeping the property of no false negatives.

Further, we give additional results following the approach in Section \ref{sec:bloomfilter:hashes} in Appendix \ref{sec:appendix:bloomfilter}.

\section{Related Work}
\label{sec:related}
The idea of learned indexes builds upon a wide range of research in machine learning and indexing techniques.
In the following, we highlight the most important related areas.

{\bf \btrees and variants:} Over the last decades a variety of different index structures have been proposed \cite{survey_btree}, such as B+-trees \cite{bplustree} for disk based systems and T-trees \cite{ttree} or balanced/red-black trees \cite{redblack1,redblack2} for in-memory systems.
As the original main-memory trees had poor cache behavior, several cache conscious B+-tree variants were proposed, such as the CSB+-tree \cite{csbtree}.
Similarly, there has been work on making use of SIMD instructions such as FAST \cite{fast} or even taking advantage of GPUs \cite{fast,hybridgpucpu,Kaczmarski2012}.
Moreover, many of these (in-memory) indexes are able to reduce their storage-needs by using offsets rather than pointers between nodes.
There exists also a vast array of research on index structures for text, such as tries/radix-trees \cite{boehmprefixtree,KISSTree,triememory}, or other exotic index structures, which combine ideas from B-Trees and tries \cite{ARTful}.

However, all of these approaches are orthogonal to the idea of learned indexes as none of them learn from the data distribution to achieve a more compact index representation or performance gains. 
At the same time, like with our hybrid indexes, it might be possible to more tightly integrate the existing hardware-conscious index strategies with learned models for further performance gains.

Since B+-trees consume significant memory, there has also been a lot of work in compressing indexes, such as prefix/suffix truncation, dictionary compression, key normalization ~\cite{survey_btree, compressing_relations_and_indexes,rdf3x}, or hybrid hot/cold indexes~\cite{reducing_storage}.
However, we presented a radical different way to compress indexes, which---dependent on the data distribution---is able to achieve orders-of-magnitude smaller indexes and faster look-up times and potentially even changes the storage complexity class  (e.g., $O(n)$ to $O(1)$ ). 
Interestingly though, some of the existing compression techniques are complimentary to our approach and could help to further improve the efficiency. 
For example, dictionary compression can be seen as a form of embedding (i.e., representing a string as a unique integer).

Probably most related to this paper are A-Trees \cite{brown_pwlf}, BF-Trees~\cite{bftree}, and B-Tree interpolation search \cite{interpolationbtree}.
BF-Trees uses a B+-tree to store information about a region of the dataset, but leaf nodes are \bloomfilters and do not approximate the CDF.
In contrast, A-Trees use piece-wise linear functions to reduce the number of leaf-nodes in a B-Tree, and \cite{interpolationbtree} proposes to use interpolation search within a B-Tree page. 
However, learned indexes go much further and propose to replace the entire index structure using learned models. 

Finally, sparse indexes like Hippo~\cite{hippo}, Block Range Indexes~\cite{postgres}, and Small Materialized Aggregates (SMAs)~\cite{sma} all store information about value ranges but again do not take advantage of the underlying properties of the data distribution.

{\bf Learning Hash Functions for ANN Indexes:}
There has been a lot of research on learning hash functions \cite{linearhashing,hash_search,hash_functions,survey_hash}.
Most notably, there has been work on learning locality-sensitive hash (LSH) functions to build Approximate Nearest Neighborhood (ANN) indexes. 
For example, \cite{nn_hash1,hash_search,hash_cnn} explore the use of  neural networks as a hash function, whereas \cite{orderhashing} even tries to preserve the order of the multi-dimensional input space. 
However, the general  goal of LSH is to group similar items into buckets to support nearest neighborhood queries, usually involving learning approximate similarity measures in high-dimensional input space using some variant of hamming distances.   
There is no direct way to adapt previous approaches to learn the fundamental data structures we consider, and it is not clear whether they can be adapted.

{\bf Perfect Hashing:}
Perfect hashing \cite{perfecthashing} is very related to our use of models for \hashmaps. 
Like our CDF models, perfect hashing tries to avoid conflicts. 
However, in all approaches of which we are aware, learning techniques have not been considered, and the size of the function grows with the size of the data.  
In contrast, learned hash functions can be independent of the size.
For example, a linear model for mapping every other integer between 0 and 200M would not create any conflicts and is independent of the size of the data. 
In addition, perfect hashing is also not useful for \btrees or \bloomfilters.

{\bf \bloomfilters:}
Finally, our existence indexes directly builds upon the existing work in \bloomfilters \cite{bloom1,bloom2}.
Yet again our  work takes a different perspective on the problem by proposing a \bloomfilter enhanced classification model or using models as special hash functions with a very different optimization goal than the hash-models we created for \hashmaps. 

{\bf Succinct Data Structures:} There exists an interesting connection between learned indexes and succinct data structures, especially rank-select dictionaries such as wavelet trees~\cite{wavelettrie,wavelettree}. 
However, many succinct data structures focus on H0 entropy (i.e., the number of bits that are necessary to encode each element in the index), whereas learned indexes try to learn the underlying data distribution to predict the position of each element. 
Thus, learned indexes might achieve a higher compression rate than H0 entropy potentially at the cost of slower operations. 
Furthermore, succinct data structures normally have to be carefully constructed for each use case, whereas learned indexes ``automate'' this process through machine learning. 
Yet, succinct data structures might provide a framework to further study learned indexes.

{\bf Modeling CDFs:} Our models for both range and point indexes are closely tied to models of the CDF.  Estimating the CDF is non-trivial and has been studied in the machine learning community  \cite{cdfmodels} with a few applications such as ranking \cite{Huang:2011:CDN:1953048.1953058}.  
How to most effectively model the CDF is still an open question worth further investigation.

{\bf Mixture of Experts:}
Our RMI architecture follows a long line of research on building experts for subsets of the data \cite{moe2}.  With the growth of neural networks, this has become more common and demonstrated increased usefulness \cite{moe}.  As we see in our setting, it nicely lets us to decouple model size and model computation, enabling more complex models that are not more expensive to execute.

\section{Conclusion and Future Work}
\label{sec:future}
We have shown that learned indexes can provide significant benefits by utilizing the distribution of data being indexed.
This opens the door to many interesting research questions.

{\bf Other ML Models:} While our focus was on linear models and neural nets with mixture of experts, there exist many other ML model types and ways to combine them with traditional data structures, which are worth exploring.

{\bf Multi-Dimensional Indexes:} Arguably the most exciting research direction for the idea of learned indexes is to extend them to multi-dimensional indexes. 
Models, especially NNs, are extremely good at capturing complex high-dimensional relationships. 
Ideally, this model would be able to estimate the position of all records filtered by any combination of attributes.

{\bf Beyond Indexing: Learned Algorithms}
Maybe surprisingly, a CDF model has also the potential to speed-up sorting and joins, not just indexes. 
For instance, the basic idea to speed-up sorting is to use an existing CDF model $F$ to put the records roughly in sorted order and then correct the nearly perfectly sorted data, for example, with insertion sort.

{\bf GPU/TPUs}
Finally, as mentioned several times throughout this paper, GPU/TPUs will make the idea of learned indexes even more valuable. 
At the same time, GPU/TPUs also have their own challenges, most importantly the high invocation latency.
While it is reasonable to assume that probably all learned indexes will fit on the GPU/TPU because of the exceptional compression ratio as shown before, it still requires 2-3 micro-seconds to invoke any operation on them.
At the same time, the integration of machine learning accelerators with the CPU is getting better \cite{nvlink,intelxeonphi} and with techniques like batching requests the cost of invocation can be amortized, so that we do not believe the invocation latency is a real obstacle.

{\bf In summary, we have demonstrated that  machine  learned models have the potential to provide significant benefits over state-of-the-art  indexes,  and we believe this is a fruitful direction for future research.}

\vspace{2mm}
{\small
\noindent {\bf Acknowledgements:} We would like to thank Michael Mitzenmacher, Chris Olston, Jonathan Bischof and many others at Google for their helpful feedback during the preparation of this paper.
}

\clearpage

\bibliographystyle{abbrv}
\bibliography{bib}

\clearpage

\appendix

\section{Theoretical Analysis of Scaling Learned Range Indexes}
\label{sec:btree:theory}
One advantage of framing learned range indexes as modeling the cumulative distribution function (CDF) of the data is that we can build on the long research literature on modeling the CDF.  Significant research has studied the relationship between a theoretical CDF $F(x)$ and the empirical CDF of data sampled from $F(x)$.  We consider the case where we have sampled i.i.d. $N$ datapoints, $\mathcal{Y}$, from some distribution, and we will use $\hat{F}_N(x)$ to denote the empirical cumulative distribution function:
\begin{align}
    \hat{F}_N(x) = \frac{\sum_{y \in \mathcal{Y}} \mathbf{1}_{y \leq x}}{N}.
\end{align}

One theoretical question about learned indexes is: how well do they scale with the size of the data $N$?
In our setting, we learn a model $F(x)$ to approximate the distribution of our data $\hat{F}_N(x)$.  Here, we assume we know the distribution $F(x)$ that generated the data and analyze the error inherent in the data being sampled from that distribution\footnote{Learning $F(x)$ can improve or worsen the error, but we take this as a reasonable assumption for some applications, such as data keyed by a random hash.}.  That is, we consider the error between the distribution of data $\hat{F}_N(x)$ and our model of the distribution $F(x)$.  
Because $\hat{F}_N(x)$ is a binomial random variable with mean $F(x)$, we find that the expected squared error between our data and our model is given by
\begin{align}
    \mathbf{E}\left[\left(F(x) - \hat{F}_N(x)\right)^2\right] = \frac{F(x)(1-F(x))}{N}.
    \label{eq:error}
\end{align}

In our application the \lookup time scales with the average error in the number of positions in the sorted data; that is, we are concerned with the error between our model $NF(x)$ and the key position $N\hat{F}_N(x)$.  With some minor manipulation of Eq. \eqref{eq:error}, we find that the average error in the predicted positions grows at a rate of $O(\sqrt{N})$.
Note that this sub-linear scaling in error for a constant-sized model is an improvement over the linear scaling achieved by a constant-sized \btree.
This provides preliminary understanding of the scalability of our approach and demonstrates how framing indexing as learning the CDF lends itself well to theoretical analysis.

\section{Separated Chaining \hashmap}
\label{appendix:chainedhash}

\begin{figure}
	\centering
    \includegraphics[width=0.6\linewidth]{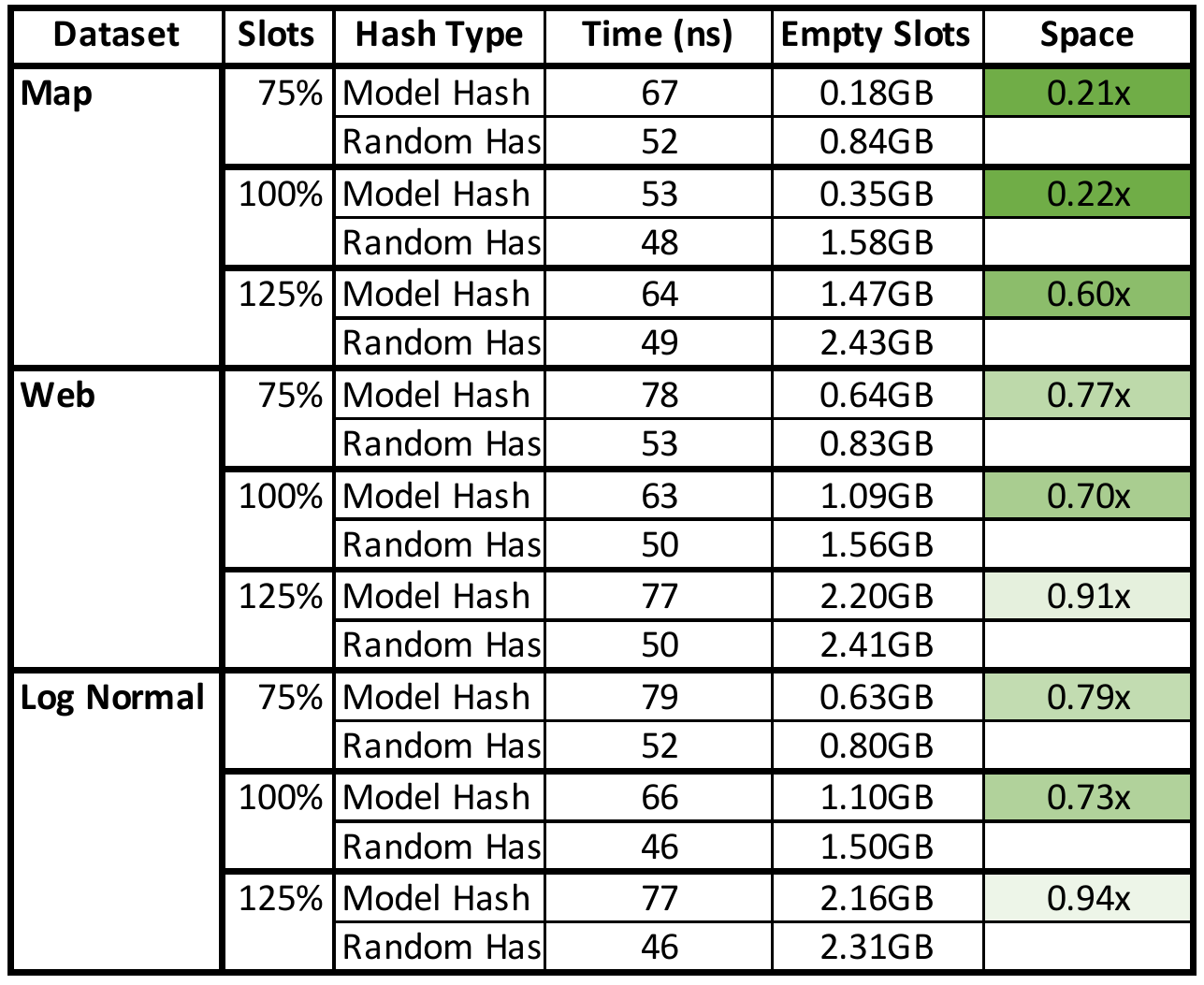}
	\caption{Model vs Random \hashmap }
     \label{fig:hash_map_results}
      \vspace{-10pt}
\end{figure}

We evaluated the potential of learned hash functions using a separate chaining \hashmap;  records are stored directly within an array and only in the case of a conflict is the record attached to the linked-list.
That is without a conflict there is at most one cache miss.
Only in the case that several keys map to the same position, additional cache-misses might occur. 
We choose that design as it leads to the best \lookup performance even for larger payloads. 
For example, we also tested a commercial-grade dense \hashmap with a bucket-based in-place overflow (i.e., the \hashmap is divided into buckets to minimize overhead and uses in-place overflow if a bucket is full \cite{googlesparsehashdoc}). 
While it is possible to achieve a lower footprint using this technique, we found that it is also twice as slow as the separate chaining approach.
Furthermore, at $80\%$ or more memory utilization the dense \hashmaps degrade further in performance. Of course many further (orthogonal) optimizations are possible and by no means do we claim that this is the most memory or CPU efficient implementation of a \hashmap.  
Rather we aim to demonstrate the general potential of learned hash functions.

As the baseline for this experiment we used our \hashmap implementation with a MurmurHash3-like  hash-function.
As the data we used the three integer datasets from Section~\ref{sec:btree:results} and as the model-based \hashmap the  2-stage RMI model with 100k models on the 2nd stage and no hidden layers from the same section. 
For all experiments we varied the number of available slots from $75\%$ to $125\%$ of the data. 
That is, with $75\%$ there are $25\%$ less slots in the \hashmap than data records. 
Forcing less slots than the data size, minimizes the empty slots within the \hashmap at the expense of longer linked lists. 
However, for \hashmaps we store the full records, which consist of a 64bit key, 64bit payload, and a 32bit meta-data field for delete flags, version nb, etc. (so a record has a fixed length of 20 Bytes); note that our chained hash-map adds another 32bit pointer, making it a 24Byte slot.  

The results are shown in Figure~\ref{fig:hash_map_results}, listing the average \lookup time, the number of empty slots in GB and the space improvement as a factor of using a randomized hash function. 
Note, that in contrast to the \btree experiments, we {\em do include the data size}. 
 The main reason is that in order to enable 1 cache-miss \lookups, the data itself has to be included in the \hashmap, whereas in the previous section we only counted the extra index overhead excluding the sorted array itself. 
 
As can be seen in Figure~\ref{fig:hash_map_results}, the index with the model hash function overall has similar performance while utilizing the memory better. 
For example, for the map dataset the model hash function only ``wastes'' 0.18GB in slots, an almost $80\%$ reduction compared to using a random hash function. 
Obviously, the moment we increase the \hashmap in size to have $25\%$ more slots, the savings are not as large, as the \hashmap is also able to better spread out the keys. 
Surprisingly if we decrease the space to $75\%$ of the number of keys, the learned \hashmap  still has an advantage because of the still prevalent birthday paradox. 

\section{Hash-Map Comparison Against Alternative Baselines}
\label{appendix:alternativehash}
In addition to the separate chaining \hashmap architecture, we also compared learned point indexes against four alternative \hashmap architectures and configurations: 

{\bf AVX Cuckoo \hashmap:} We used an AVX optimized Cuckoo \hashmap from \cite{dawngit}. 

{\bf Commercial Cuckoo \hashmap:} The implementation of \cite{dawngit} is highly tuned, but does not handle all corner cases. We therefore also compared against a commercially used Cuckoo \hashmap. 

{\bf In-place chained \hashmap with learned hash functions:} One significant downside of separate chaining is that it requires additional memory for the linked list. As an alternative, we implemented a chained \hashmap, which uses a two pass algorithm: in the first pass, the learned hash function is used to put items into slots. If a slot is already taken, the item is skipped. Afterwards we use a separate chaining approach for every skipped item except that we use the remaining free slots with offsets as pointers for them. 
As a result, the utilization can be 100\% (recall, we do not consider inserts) and the quality of the learned hash function can only make an impact on the performance not the size: the fewer conflicts, the fewer cache misses. 
We used a  simple single stage multi-variate model as the learned hash function and implemented the \hashmap including the model outside of our benchmarking framework to ensure a fair comparison. 

\begin{table}[h!]
\centering
 \begin{tabular}{| p{10cm} | p{1.5cm}  | c|} 
 \hline
  Type & Time (ns) & Utilization \\ 
 \hline\hline
 AVX Cuckoo, 32-bit value & 31ns & 99\%  \\ 
 \hline
 AVX Cuckoo, 20 Byte record& 43ns & 99\%  \\
 \hline
 Comm. Cuckoo, 20Byte record & 90ns & 95\%  \\
 \hline \hline
 In-place chained \hashmap with learned hash functions, record & 35ns & 100\%  \\
 \hline
\end{tabular}
\caption{\hashmap alternative baselines}

\label{table:hashbaselines}
\end{table}

Like in Section~\ref{appendix:chainedhash} our  records are 20 Bytes large and consist of a 64bit key, 64bit payload, and a 32bit meta-data field as commonly found in real applications (e.g., for delete flags, version numbers, etc.). 
For all \hashmap architectures we tried to maximize utilization and used records, except for the AVX Cuckoo \hashmap where we also measured the performance for 32bit values. 
As the dataset we used the log-normal data and the same hardware as before. 
The results  are shown in Table~\ref{table:hashbaselines}. 

The results for the AVX cuckoo \hashmap show that the payload has a significant impact on the performance. 
Going from 8 Byte to 20 Byte decreases the performance by almost 40\%. 
Furthermore, the commercial implementation which handles all corner cases but is not very AVX optimized slows down the lookup  by another factor of 2. 
In contrast, our learned hash functions with in-place chaining can provide better lookup performance than even the cuckoo \hashmap for our records. 
The main take-aways from this experiment is that learned hash functions can be used with different \hashmap architectures and that the benefits and disadvantages highly depend on the implementation, data and workload.

\section{Future Directions for Learned \btrees}
\label{appendix:btree:future}
In the main part of the paper, we have focused on index-structures for read-only, in-memory  database systems.
Here we outline how the idea of learned index structures could be extended in the future.

\subsection{Inserts and Updates}
\label{sec:rmi:insers}
On first sight, inserts seem to be the Achilles heel of learned indexes because of the potentially high cost for learning models, but yet again learned indexes might have a significant advantage for certain workloads. 
In general we can distinguish between two types of inserts: (1) {\em appends} and (2) {\em inserts in the middle} like updating a secondary index on the customer-id over an order table.  

Let's for the moment focus on the first case: appends. 
For example, it is reasonable to assume that for an index over the timestamps of web-logs, like in our previous experiments, most if not all inserts will be appends with increasing timestamps. 
Now, let us further assume that our model generalizes and is able to learn the patterns, which also hold for the future data.
As a result, updating the index structure becomes an $O(1)$ operation; it is  a simple append and no change of the model itself is needed, whereas a \btree requires $O(\log n)$ operations to keep the \btree balance. 
A similar argument can also be made for inserts in the middle, however, those might require to move data or reserve space within the data, so that the new items can be put into the right place. 

Obviously, this observation also raises several questions. 
First, there seems to be an interesting trade-off in the generalizability of the model and the ``last mile'' performance; the better the ``last mile'' prediction, arguably, the more the model is overfitting and less able to generalize to new data items. 

Second, what happens if the distribution changes? Can it be detected, and is it possible to provide similar strong guarantees as \btrees which always guarantee $O(log n)$ \lookup and insertion costs? 
While answering this question goes beyond the scope of this paper, we believe that it is possible for certain models to achieve it.
More importantly though, machine learning offers new ways to adapt the models to changes in the data distribution, such as online learning, which might be  more effective than traditional \btree balancing techniques. 
Exploring them also remains future work. 

Finally, it should be pointed out that there always exists a much simpler alternative to handling inserts by building a delta-index \cite{delta-index}.
All inserts are kept in buffer and from time to time merged with a potential retraining of the model.
This approach is already widely used, for example in Bigtable \cite{bigtable} and many other systems, and was recently explored in \cite{brown_pwlf} for learned indexes.

\subsection{Paging}
\label{sec:rmi:paging}
Throughout this section we assumed that the data, either the actual records or the  \texttt{<key,pointer>} pairs, are stored in one continuous block. 
However, especially for indexes over data stored on disk, it is quite common to partition the data into larger pages that are stored in separate regions on disk.
To that end, our observation that a model learns the CDF no longer holds true as $\textit{pos} = {\rm Pr}(X<\text{Key}) * N$ is violated.
In the following we outline several options to overcome this issue:

Leveraging the RMI structure: The RMI structure already partitions the space into regions. 
With small modifications to the learning process, we can minimize how much models overlap in the regions they cover.
Furthermore, it might be possible to duplicate any records which might be accessed by more than one model. 

Another option is to have an additional translation table in the form of \texttt{<first_key, disk-position>}. 
With the translation table the rest of the index structure remains the same. 
However, this idea will  work best if the disk pages are very large.
At the same time it is possible to use the predicted position with the min- and max-error to reduce the number of bytes which have to be read from a large page, so that the impact of the page size might be negligible. 

With more complex models, it might actually be possible to learn the actual pointers of the pages. Especially if a file-system is used to determine the page on disk with a systematic numbering of the blocks on disk (e.g., \texttt{block1,...,block100}) the learning process can remain the same.

Obviously, more investigation is required to better understand the impact of learned indexes for disk-based systems. At the same time the significant space savings as well as speed benefits make it a very interesting avenue for future work.

\section{Further \bloomfilter Results}
\label{sec:appendix:bloomfilter}
In Section \ref{sec:bloomfilter:hashes}, we propose an alternative approach to a learned \bloomfilter where the classifier output is discretized and used as an additional hash function in the traditional Bloom filter setup.  Preliminary results demonstrate that this approach in some cases outperforms the results listed in Section \ref{sec:bloomfilter:results}, but as the results depend on the discretization scheme, further analysis is worthwhile.  We describe below these additional experiments.

As before, we assume we have a model model $f(x) \rightarrow [0,1]$ that maps keys to the range $[0,1]$.  
In this case, we allocate $m$ bits for a bitmap $M$ where we set $M[\lfloor m f(x) \rfloor] = 1$ for all inserted keys $x \in \mathcal{K}$.  
We can then observe the FPR by observing what percentage of non-keys in the validation set map to a location in the bitmap with a value of 1, i.e. ${\rm FPR}_m \equiv \frac{\sum_{x \in \mathcal{\tilde{U}}} M[\lfloor f(x)m\rfloor] }{|\mathcal{\tilde{U}}|}$.  
In addition, we have a traditional Bloom filter with false positive rate ${\rm FPR}_B$.  We say that a query $q$ is predicted to be a key if $M[\lfloor f(q)m\rfloor] = 1$ and the Bloom filter also returns that it is a key. 
As such, the overall FPR of the system is ${\rm FPR}_m \times {\rm FPR}_B$; we can determine the size of the traditional Bloom filter based on it's false positive rate ${\rm FPR}_B = \frac{p^*}{ {\rm FPR}_m}$ where $p^*$ is the desired FPR for the whole system.

As in Section \ref{sec:bloomfilter:results}, we test our learned \bloomfilter on data from Google's transparency report.  We use the same character RNN trained with a 16-dimensional width and 32-dimensional character embeddings.  Scanning over different values for $m$, we can observe the total size of the model, bitmap for the learned Bloom filter, and the traditional Bloom filter.  For a desired total FPR $p^* = 0.1\%$, we find that setting $m=1000000$ gives a total size of 2.21MB, a 27.4\% reduction in memory, compared to the 15\% reduction following the approach in Section \ref{sec:bloomfilter:classification} and reported in Section \ref{sec:bloomfilter:results}.  For a desired total FPR $p^* = 1\%$ we get a total size of 1.19MB, a 41\% reduction in memory, compared to the 36\% reduction reported in Section \ref{sec:bloomfilter:results}.

These results are a significant improvement over those shown in Section \ref{sec:bloomfilter:results}.  However, typical measures of accuracy or calibration do not match this discretization procedure, and as such further analysis would be valuable to understand how well model accuracy aligns with it's suitability as a hash function.

\end{document}